\documentclass[a4paper,11pt]{article}
\pdfoutput=1
\usepackage{jheppub}
\usepackage{amsmath}
\usepackage{slashed}
\usepackage{amssymb}
\usepackage{color}
\usepackage{graphicx}
\usepackage{hyperref}
\usepackage{xspace,slashed}
\usepackage[utf8]{inputenc}
\usepackage{subcaption}
\usepackage{multirow}
\usepackage{algorithm}
\usepackage{algorithmic}
\usepackage{stackengine}
\usepackage{enumitem}
\usepackage{etoolbox}

\usepackage{tikz}
\usepackage{tikz-feynman}

\newtoggle{draft}

\toggletrue{draft}
\iftoggle{draft} {
  \usepackage{changes}
}{%
\usepackage[final]{changes}
}

\definechangesauthor[name=pn, color=blue]{pn}
\definechangesauthor[name=km, color=magenta]{km}
\definechangesauthor[name=mao, color=red]{mao}

\usepackage{booktabs}

\makeatletter
\g@addto@macro\bfseries{\boldmath}
\makeatother

\newcommand{\barhatN}{{\bar{\hatN}}}

\newcommand{\hatone}{{\bf 1}}
\newcommand{\hatN}{{\bf N}}
\newcommand{\hatS}{{\bf S}}
\newcommand{\hatC}{{\bf \Sigma}}
\newcommand{\ptop}{p_t}
\newcommand{\ptopbar}{p_{\bar{t}}}
\newcommand{\hatp}{{\slashed p}}

\newcommand{\tildeptop}{{\tilde p}_t}
\newcommand{\tildeptopbar}{{\tilde p}_{\bar{t}}}

\newcommand{\mt}{m_t}
\newcommand{\tildemt}{{\tilde m}_t}
\newcommand{\qtop}{q_t}

\newcommand{\hatk}{{\slashed k}}

\newcommand{\bt}{{\bar t}}

\newcommand{\as}{\alpha_s}
\newcommand{\gs}{g_s}
\newcommand\nf{n_{f}}

\newcommand\Ca{C_A}
\newcommand\Cf{C_F}

\newcommand{\be}{\begin{equation}}
\newcommand{\ee}{\end{equation}}

\newcommand{\xa}{\mathfrak{a}}

\newcommand{\Nset}{N}
\newcommand{\qset}{N_q}
\newcommand{\qbarset}{N_{\bar q}}

\usepackage{soul}
\allowdisplaybreaks

\definecolor{azure}{rgb}{0.0, 0.9, 1.0}

\newcommand{\TTPaff}{Institute for Theoretical Particle Physics,
  KIT, 76128 Karlsruhe, Germany}
\newcommand{\MILaff}{INFN, Sezione di Milano-Bicocca, and Universit\`a di Milano-Bicocca, Piazza della Scienza 3, 20126 Milano, Italy}

\newcommand{\Saclay}{Universit\'e Paris-Saclay, CNRS, IJCLab, 91405 Orsay, France}

\preprint{
  \begin {flushright}
    TTP23-032, P3H-23-055
  \end{flushright}
}
\title{
Linear power corrections to top quark pair production in hadron collisions
}
\author[a]{Sergei Makarov,}
\author[a]{Kirill Melnikov,}
\author[b]{Paolo Nason,}
\author[c]{Melih A. Ozcelik}
\affiliation[a]{\TTPaff}
\affiliation[b]{\MILaff}
\affiliation[c]{\Saclay}
\emailAdd{sergei.makarov@kit.edu}
\emailAdd{kirill.melnikov@kit.edu}
\emailAdd{paolo.nason@mib.infn.it}
\emailAdd{melih.ozcelik@ijclab.in2p3.fr}

\abstract{ We compute, in the framework of renormalon calculus, the 
${\cal O}(\Lambda_{\rm QCD})$ 
corrections to the production of $t\bar{t}$ pairs in hadron collisions under the assumption that $q \bar q \to t \bar t$ is the dominant partonic channel. This assumption is not applicable to top quark pair production at the LHC but it is valid for the Tevatron where collisions of 
protons and anti-protons 
were studied.  We show that the linear power correction to the total $t \bar t$ production cross section vanishes provided one uses a short-distance scheme for the top quark mass.
We also derive relatively simple formulas for the power corrections to top quark kinematic distributions. Although small numerically, these power corrections exhibit interesting dependencies on top quark kinematics. 
 }

\begin{document}

\maketitle 
%==
\section{Introduction}

Top quark studies play a central role in the current exploration of the Standard Model of particle physics and in the quest to discover physics beyond it. Of particular interest are top quark couplings to electroweak gauge bosons and the Higgs boson, as well as its mass and width \cite{ATLAS:2018mme,ATLAS:2020ior}. It is well-known that the lifetime of the top quark is so short that the hadronisation mechanism has no time to set in. As a result, many properties of ``free'' top quarks, such as their polarisations and spin correlations, can be accessed by studying kinematic distributions of their decay products.

Given the very rich research  program that can be pursued by studying top quarks, experimental and theoretical exploration  of top quark pair production progressed rapidly in recent years and reached a very advanced stage \cite{Mazzitelli:2021mmm, ATLAS:2020aln,CMS:2020cga,Czakon:2013goa,Czakon:2016ckf,Catani:2020tko,Kulesza:2020nfh,Kidonakis:2022qvz,Bevilacqua:2022ozv,Denner:2023eti}.   In fact, progress in theory and experiment allows us to study various properties  of top quarks with very high precision making a better understanding of subtle effects desirable and even  mandatory in certain cases. 

One important class of such effects are the non-perturbative power corrections. In the case of top quark pair production, these power corrections are especially important for the extraction of the top quark mass from the total cross section and from kinematic distributions \cite{ATLAS:2014nxi,CMS:2016yys,CMS:2018fks,ATLAS:2019hau,ATLAS:2017dhr,CMS:2019esx,CMS:2014rml,Juste:2013dsa,Hoang:2014oea,Nason:2017cxd,Azzi:2019yne}. Since a sound theoretical understanding of power corrections to top quark pair production is lacking,\footnote{We note that power corrections to processes and observables  amenable to the operator product expansion can be characterised in terms of  expectation values of higher-dimensional operators.  Unfortunately, understanding collider processes in such a framework is an open problem.
} it cannot be excluded that such extractions are biased.  

A valuable approach to the study of power corrections is  the renormalon calculus in the  large--$b_0$ approximation. It can be applied to processes that, at the Born level, are described by Feynman diagrams without gluons. The method consists  in adding  one soft gluon (virtual or real), dressed with an arbitrary number of quark-anti-quark vacuum 
polarisation insertions, to the Born process. The underlying
abelianised model corresponds to QCD in the limit of a large, negative number of quark flavours. In this limit, the theory remains asymptotically free, and exhibits infrared renormalons. It turns out that the results in the large-$b_0$ approximation can be easily obtained from calculations in QCD where the gluon carries a small mass $\lambda$. In particular, ${\cal O}(\Lambda_{\rm QCD})$ corrections are associated with corrections of order $\lambda$ to the considered  observables. This procedure is well known; for example, it is reviewed in ref.~\cite{Beneke:1998ui}, where many applications are illustrated. Furthermore,  a comprehensive  account of the method is given in Appendix~B of ref.~\cite{FerrarioRavasio:2018ubr}.

In two recent papers~\cite{Caola:2021kzt,Makarov:2023ttq}, some of us have used the renormalon calculus to discuss  linear power corrections to some collider observables.\footnote{
Other approaches to understanding power corrections to hard processes at lepton and hadron colliders   are discussed in  refs.~\cite{Dehnadi:2023msm,Bachu:2020nqn,Hoang:2017kmk,Hoang:2019ceu}.} In particular, in ref.~\cite{Makarov:2023ttq}, the case of $t$-channel single top production was considered. This  process does not have any gluon at the Born level and is thus amenable to  renormalon calculus. It was found that no linear power corrections are present  in the total inclusive cross section 
of 
the $t$-channel single top production, provided that the cross section is expressed in terms of a short-distance mass of the top quark. We note in passing that, in general, certain input parameters, such as   e.g. masses of heavy quarks in the on-shell renormalisation scheme, may receive linear power corrections, and it is better to   avoid them  by switching to input parameters defined at short distances.  However, employing  the short-distance quark masses is not sufficient to get rid of linear power corrections in general since, as  it was shown in ref.~\cite{Makarov:2023ttq}, they exist in  top-quark kinematic distributions. Such corrections are easily calculable within the same renormalon framework that we use to study total inclusive cross sections.

The goal of this paper is to go one step further in the computation of the linear power corrections to top quark production in hadron collisions, and 
to use the renormalon calculus to study such corrections in the $q \bar q \to t \bar t$ partonic  channel. We note that this process  is mediated by a gluon exchange at leading order and that  such a gluon is highly virtual.
As we explain below, the large virtuality of the gluon allows us to use the  Low-Burnett-Kroll (LBK) theorem \cite{Low:1958sn,Burnett:1967km},\footnote{For recent literature on the LBK theorem see ref.~\cite{Engel:2021ccn} and references therein.} to uniquely reconstruct the first subleading term in the soft expansion of the $q \bar q \to t \bar t$ amplitude, and in this way compute the linear power correction.
On the contrary, the $gg \to t \bar t$ channel, which is dominant at the LHC, has  on-shell gluons as external lines, and we will not deal with it in the present work.

The remainder of the paper is organised as follows. In Section~\ref{sect:2} we describe the generalisation of the Low-Burnett-Kroll theorem to processes with arbitrary number of quarks and anti-quarks as external particles.  In Section~\ref{sect:3} we discuss aspects of the large-$N_f$ limit of QCD which concern  the presence of a virtual gluon in the Born amplitude. 
In Section~\ref{sec:cancellation} we explore the structure of cancellations of various ${\cal O}(\lambda)$ terms and show that they occur  \emph{independently} for different colour dipoles responsible for  soft QCD radiation and for contributions where the same parton emits and absorbs soft radiation.  
In Section~\ref{sect:kinematics} we continue with the discussion of ${\cal O}(\lambda)$ corrections to top-quark kinematic distributions.
We apply these general results to the partonic process
$ q \bar q \to t \bar t$  in Section~\ref{sec:applications}, where we also compute 
non-perturbative corrections to  the 
top quark pair production in $ p \bar p$  collisions. 
We conclude in Section~\ref{sect:concl}. The Appendices 
contain useful technical information and  
results for non-perturbative corrections to observables in a general $q \bar q \to t \bar t +X$ process. 

\section{The Low-Burnett-Kroll theorem}
\label{sect:2}

The goal of this section is to discuss the Low-Burnett-Kroll theorem for processes with an arbitrary number of external quarks and anti-quarks that carry non-Abelian charges, and an arbitrary number of colour-neutral particles. We consider the process 
\begin{equation}
  \varnothing \to \sum \limits_{i=1}^{N}  f_i(p_i) + X(p_X),
  \label{eq1.1a}
\end{equation}
where $f_i(p_i)$ is a quark or an anti-quark of  flavour $i$ with  momentum $p_i$ and mass $m_i$, and  $X(p_X)$ denotes, collectively, a system of colour-neutral
particles, and the symbol $\varnothing$ in the initial state symbolises the vacuum. We only consider final-state particles, since amplitudes with initial-state particles  can be obtained from our results by crossing. The total number of colour-charged particles is $N$ and we will use $N_q$ and $N_{\bar q}$ to refer
to the total number of quarks and anti-quarks, respectively.\footnote{Obviously, colour conservation requires $N_q = N_{\bar q}$.} With a slight abuse of notation we will also indicate with $\Nset$, $\qset$ and $\qbarset$ the sets of all quark and anti-quark indices. We stress that gluons cannot appear as external on-shell particles but virtual gluons can be present as internal lines in the Born amplitude.

\subsection{Real emission contribution}

We consider the emission of a gluon with momentum $k$ in the process shown in eq.~(\ref{eq1.1a}),
\begin{equation}
  \varnothing \to \sum \limits_{i=1}^{N} f_i(p_i) + X(p_X) + g(k).
  \label{eq1.1b}
\end{equation}
This gluon is considered to be massive, with a tiny mass $\lambda$. We are interested in a situation when the emitted gluon is soft and has an energy comparable to its mass.

\begin{figure}
    \centering
    \begin{subfigure}[t]{0.32\textwidth}
        \includegraphics[width=1\textwidth]{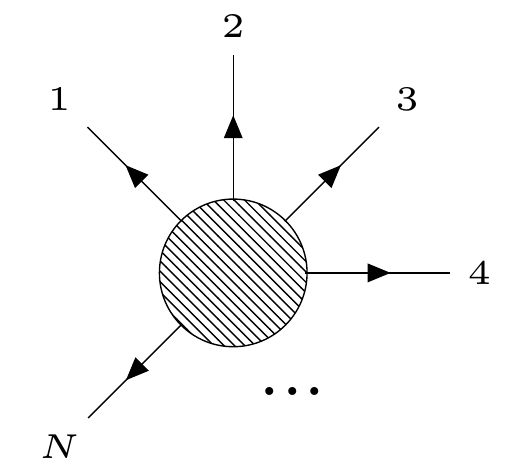}
    \end{subfigure}
    \hfill
    \begin{subfigure}[t]{0.32\textwidth}
        \raisebox{0pt}{\includegraphics[width=1.12\textwidth]{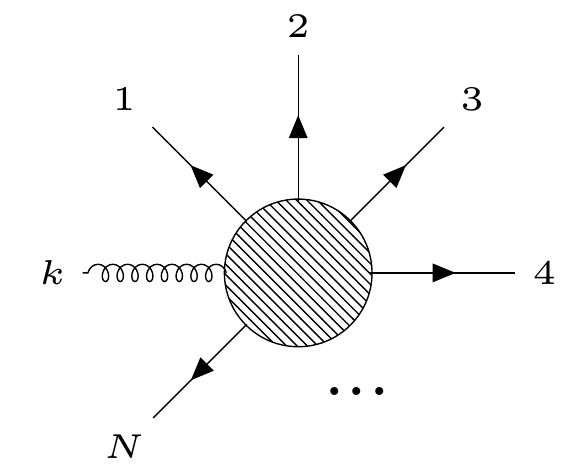}}
    \end{subfigure}
    \hfill
    \begin{subfigure}[t]{0.32\textwidth}
        \includegraphics[width=1\textwidth]{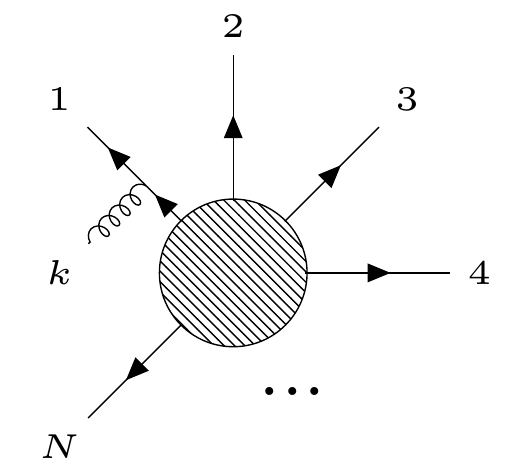}
    \end{subfigure}
    \caption{Leading order and the relevant real-emission contributions to the process $\varnothing \to \sum_i f_i(p_i) + X(p_X) + g(k)$. Arrows show directions of outgoing momenta for both quarks and anti-quarks.
    }
    \label{fig1}
\end{figure}

We extract  the gluon polarisation vector $\epsilon$ and   write the amplitude of the process in eq.~\eqref{eq1.1b}
as 
\begin{equation}
  {\cal A}_{\rm real} = \epsilon_\mu \langle c| {\cal M}^{a,\mu} \rangle, 
\end{equation}
where $|c \rangle $ indicates colour quantum numbers of all particles in eq.~(\ref{eq1.1a}) 
and $a$ is the gluon colour index. Thus ${\cal M}^{a,\mu}$ is a vector in colour space,
in a representation that is the direct product of fundamental or anti-fundamental representations, each one associated to an outgoing particle or anti-particle. In this context we also define the colour space operator $T^a_i$, corresponding to a Gell-Mann matrix acting
on the vector subspace of the outgoing particle $i$. For an outgoing anti-particle it is minus the transpose of the Gell-Mann matrix that acts by left multiplication. It is more convenient to us to drop the minus sign, and have instead the Gell-Mann matrix $T^a_i$ act by right multiplication upon the colour index of the outgoing anti-particle.

Separating the emissions from the external legs and the ``structure-dependent'' radiation from the internal lines as shown in Fig.~\ref{fig1}, we write the reduced amplitude ${\cal M}^{a,\mu}$ as\footnote{We will write ${\cal M}^{a,\mu}$ rather than 
$|{\cal M}^{a,\mu} \rangle$ in what follows to simplify the 
notations.}
\begin{equation}
\begin{split} 
  \label{eq8.3}
      {\cal M}^{a,\mu} & = g_s \sum_{i \in \qset} \bar{u}(p_i) \gamma^\mu \frac{\hatp_i + \hatk + m_i}{d_i} \; T^a_i \; \hatN_i(p_1,p_2,..,p_i + k,...)
\\
      & + g_s \sum \limits_{i \in \qbarset} \hatN_i(p_1,p_2,..,p_i + k,...)  \; T^a_i \frac{-\hatp_i - \hatk + m_i}{d_i} \gamma^\mu v(p_i)
+ {\cal M}^{a,\mu}_{\rm reg}(p_1,..,p_N; k).
\end{split} 
\end{equation}
Here, $d_i = (p_i+k)^2 - m_i^2$, $T_i^a$ refers to the colour charge of parton $i$ and $\hatN_i(p_1,p_2,..,p_i + k,...)$ is the Green's function of the process in eq.~(\ref{eq1.1a})  with an amputated off-shell leg $i$. The amplitude ${\cal M}^{a,\mu}_{\rm reg}$ describes the structure-dependent radiation; it is regular in the  $k \sim \lambda \to 0$ limit. 

The LBK theorem stems from the observation that one can determine ${\cal M}^{a,\mu}_{\rm reg}$ at $k = 0$ by requiring that $k_\mu {\cal M}^{a,\mu}=0$.\footnote{
We note that recently  the validity of LBK theorem has been questioned in refs.~\cite{Lebiedowicz:2021byo,Lebiedowicz:2023mlz,Lebiedowicz:2023ell}. Although some of the criticism in these references 
might be justified, we believe that  our derivation 
of the theorem is consistent and leads 
to correct results, see  Section~\ref{sec:validity} for further discussion of this point.} 
We will apply this observation to eq.~\eqref{eq8.3}. Before we do that, it is convenient to rewrite the parts of the amplitude that describe the gluon emissions from the external legs. Using the Dirac equation, we obtain
\be
\bar u_i \gamma^\mu \; \frac{\hatp_i + \hatk + m_i}{d_i} =
\bar u_i \left ( J_i^\mu + \hatS_i^\mu \right ),
\ee
where
\be
J_i^\mu = \frac{2 p^\mu_i + k^\mu}{d_i},\;\;\;\; \hatS_i^\mu = \frac{\sigma^{\mu \nu} k_\nu}{d_i},
\label{eq2.6}
\ee
with $\sigma^{\mu \nu} = 1/2\; [ \gamma^\mu, \gamma^\nu]$. We note that these quantities have the following properties
\be
k_\mu J_i^\mu =1,\;\;\;\;\;\; k_\mu \hatS_i^\mu = 0.
\label{eq2.7}
\ee
For an anti-quark, we find a similar equation 
\be
 \frac{-\hatp_i - \hatk + m_i}{d_i} \gamma^\mu  v_i =
 \left (- J_i^\mu + \hatS_i^\mu \right )  v_i.
\ee

We contract eq.~\eqref{eq8.3} with $k_\mu$, use eq.~\eqref{eq2.7} and obtain
\be \label{eq:Mcontr}
0 = g_s \sum \limits_{i=1}^{N} \eta_{i} {\cal N}^a_i(p_1,..,p_i+k,...) + k_\mu {\cal M}^{a,\mu}_{\rm reg}(p_1,..,p_N; k),
\ee
where  we introduced
\be \label{eq:Nsymbols}
{\cal N}_i^a = \bar{u}_i T_i^a \hatN_i ~~~{\rm or} \;\;\;\;  {\cal N}_i^a = \hatN_i T_i^a v_i,
\ee
as appropriate for a quark or an anti-quark,  and $\eta_i = 1$ or $-1$ if $i$ is a quark or an anti-quark. 

Expanding equation (\ref{eq:Mcontr}) in Taylor series through linear terms in $k$, we obtain\footnote{Note that our conventions for the colour generators differ from the ones  commonly used in the literature. In this paper, we use $(T_i)^a_{\alpha \beta}=t^a_{\alpha \beta}$ both for outgoing particles
and anti-particles, except that in the latter case $(T_i)^a_{\alpha \beta}$ acts on the right, corresponding to the transposed matrix acting on the left.
Besides transposing, the anti-fundamental representation requires
a minus sign, that we absorb into the factors $\eta_i$.}
\be
\begin{split} 
  & 0 =  \sum \limits_{i=1}^{N} \eta_i \; {\cal  N}^a_i(p_1,..,p_i,..,p_N),
  \\
  & 0 =  k^\mu \left ( g_s \sum \limits_{i=1}^{N}  \eta_i  {\bar D}_{i,\mu} {\cal N}^a_i(p_1,..,p_i,..,p_N) +  {\cal M}^a_{\rm reg,\mu}(p_1,..,p_N; 0) \right ),
\end{split}
\label{eq2.11}
\ee
where $D_{i,\mu} = \partial/\partial p_{i}^\mu$ and $\bar D$ indicates that the differential operator does not act
on the external spinors that appear in ${\cal N}_i^a$ defined in  eq.~(\ref{eq:Nsymbols}).

We also note that at $k = 0$ the functions ${\cal  N}^a_i$ can  be written as 
\be
{\cal  N}^a_i = T^a_{i} | {\cal M}_0(p_1,p_2,..,p_N) \rangle ,
\ee
where  ${\cal M}_0(p_1,p_2,..,p_N) $ is the amplitude of the process in eq.~(\ref{eq1.1a}) and we have written it as
a vector in colour space. 

The first equation in eq.~(\ref{eq2.11}) is the colour conservation condition. The second equation has to be satisfied for arbitrary $k$ so that
\be
| {\cal M}^{a,\mu}_{\rm reg}(p_1,..,p_N; 0) \rangle
= - g_s \sum \limits_{i=1}^{N} \eta_i {\bar D}_{i}^{\mu} {\cal N}^a_i
 = -g_s \sum \limits_{i=1}^{N} \eta_i {\bar D}_{i}^{\mu} T_i^a | {\cal M}_0 \rangle .
   \label{eq2.12}
\ee

Having determined the structure-dependent part of the real-emission amplitude, we can now write the full amplitude
as an expansion in the gluon momentum with ${\cal O}(k^0)$ accuracy. We obtain
\be
   {\cal M}^\mu =
   g_s \sum \limits_{i \in \Nset} \eta_i ( J_i^\mu + {\bar L}_i^\mu) T_i^a \; |{\cal M}_0 \rangle 
   + g_s \sum \limits_{i \in \qset} \bar u_i \hatS_i^\mu  \hatN_i^a
   + g_s \sum \limits_{i \in \qbarset}     \hatN_i^a \hatS_i^\mu v_i
   + {\cal O}(k),
   \label{eq2.13}
\ee
where
\be
\hatN_i^a = T_i^a \hatN_i~~~{\rm or}~~~\hatN_i T_i^a,
\ee
depending on whether parton $i$ is a quark or an anti-quark,  and 
\be\label{eq:Ldef}
{\bar L}_i^\mu = J_i^\mu k^\nu {\bar D}_{i,\nu} - {\bar D}_{i}^\mu.
\ee

As the next step, we need to compute the square of the gluon emission amplitude summed  over polarisations and colours
of external particles.  Working through the first subleading term in the expansion of the gluon momentum, we find
\be
\begin{split} 
   {\cal M}_\mu^{\dagger} {\cal M}^\mu
   =& g_s^2 \sum \limits_{i,j \in \Nset}
     \eta_i \eta_j  \langle {\cal M}_0 | J_i^\mu J_{j,\mu} T_i^a T_j^a 
     +    \overset{\leftarrow}{{\bar L}}_{j,\mu}  J_i^\mu  T_j^a T_i^a 
     +  J_j^\mu   T_j^a T_i^a {\bar L}_{i,\mu}| {\cal M}_0 \rangle
             \\
             + &g_s^2 \sum \limits_{i,j \in \Nset } \eta_i J_{i,\mu} \left (
             \langle {\cal M}_{0,j}^{s,\mu}| T_j^a T_i^a | {\cal M}_0 \rangle
             + \langle {\cal M}_0 | T_i^a T_j^a | {\cal M}_{0,j}^{s,\mu} \rangle \right ),
   \end{split} 
\label{eq2.16}
\ee
where
\be
   | {\cal M}_{0,j}^{s,\mu} \rangle  =
   \left \{
\begin{array}{cc}
    \bar u_j \hatS_{j}^{\mu} \hatN_j, & j  \in \qset, \\
      \hatN_j  \hatS_{j}^{\mu} v_j , & j  \in \qbarset.
\end{array}
\right.
\label{eq:strucdependM0}
\ee
We note that
\be
   \langle  {\cal M}_{0,j}^{s,\mu} |  =
   (-1) \left \{
\begin{array}{cc}
      \bar {\hatN}_j  \hatS_{j}^{\mu} u_j,   & j  \in \qset, \\
      \bar v_j \hatS_{j}^{\mu} \bar {\hatN}_j, & j  \in \qbarset,
\end{array}
\right.
\ee
because the spin operators  defined in eq.~(\ref{eq2.6}) are anti-hermitian.

We now discuss the various terms that appear 
in eq.~(\ref{eq2.16}). The terms in the first line can be easily simplified if we use the fact that $ T_j^a T_i^a = T_i^a T_j^a$. Then we find
\be
\begin{split}
& \sum \limits_{i,j \in \Nset}
     \eta_i \eta_j  \langle {\cal M}_0 | J_i^\mu J_{j,\mu} T_i^a T_j^a 
     +    \overset{\leftarrow}{{\bar L}}_{j,\mu}  J_i^\mu  T_j^a T_i^a 
     +  J_j^\mu   T_j^a T_i^a {\bar L}_{i,\mu}| {\cal M}_0 \rangle
     \\
     & =
     \sum \limits_{i,j \in \Nset}
     \eta_i \eta_j \left ( J_i^\mu J_{j,\mu} + J_i^\mu {\bar L}_{j,\mu} \right ) F^{ij}_{\rm LO},
  \end{split}
\label{eq2.17}
\ee
where
\be
F^{ij}_{\rm LO} = \langle {\cal M}_0 | T^a_i T^a_j | {\cal M}_0 \rangle,
\ee
is the colour-correlated matrix element squared of the
process in eq.~(\ref{eq1.1a}).

Next, we need to consider the various contributions that depend on the spin operators $\hatS_i^\mu$. As we will see, in this case one should be careful with the relative signs between the quark and the anti-quark cases. 
Consider the expression
\be
        \sum \limits_{i \in \Nset, j \in \qset } \eta_i J_i^\mu \left (
             \langle {\cal M}_{0,j}^{s,\mu}| T_j^a T_i^a | {\cal M}_0 \rangle
             + \langle {\cal M}_0 | T_i^a T_j^a | {\cal M}_{0,j}^{s,\mu} \rangle \right ).
\ee
To simplify it, we note that we can write a tree-level amplitude as
\be
| {\cal M}_{0} \rangle  = \bar u_j \hatN_j,
\ee
``factoring out'' a spinor $u_j$. Then we find 
\be
\begin{split}
   &     \sum \limits_{i \in \Nset, j \in \qset } \eta_i J_i^\mu \left (
             \langle {\cal M}_{0,j}^{s,\mu}| T_j^a T_i^a | {\cal M}_0 \rangle
             + \langle {\cal M}_0 | T_i^a T_j^a | {\cal M}_{0,j}^{s,\mu} \rangle \right )
             \\
  & =  \sum \limits_{i \in \Nset, j \in \qset } \eta_i J_i^\mu \left ( -\bar u_j \hatN_j T_i^a T_j^a \bar{\hatN}_j \hatS_{j,\mu} u_j
+  \bar u_j \hatS_{j,\mu} \hatN_j  T_i^a T_j^a \bar{\hatN}_j u_j \right )
\\
& = \sum \limits_{i \in \Nset,j \in \qset} \eta_i J_i^\mu {\rm Tr} \left (
 [\hat \rho_{q,j}, \hatS_{j,\mu} ] \hatN_j T_i^a T_j^a  \bar{\hatN}_j
 \right ),
 \label{eq2.20}
 \end{split} 
\ee
where  $\hat \rho_{q,j} = \hatp_j + m_j$ is the density matrix associated with the quark  $j$. A similar calculation
for an anti-quark gives
\be
\begin{split}
   &     \sum \limits_{i \in \Nset, j \in \qbarset } \eta_i J_i^\mu \left (
             \langle {\cal M}_{0,j}^{s,\mu}| T_j^a T_i^a | {\cal M}_0 \rangle
             + \langle {\cal M}_0 | T_i^a T_j^a | {\cal M}_{0,j}^{s,\mu} \rangle \right )
             \\
  & =  \sum \limits_{i \in \Nset, j \in \qbarset } \eta_i J_i^\mu \left ( -\bar v_j \hatS_{j,\mu} \bar{\hatN}_j T_i^a T_j^a \hatN_j v_j
+  \bar v_j  \bar{\hatN}_j  T_i^a T_j^a {\hatN}_j \hatS_{j,\mu} v_j \right )
\\
& = -\sum \limits_{i \in \Nset,j \in \qbarset} \eta_i J_i^\mu {\rm Tr} \left (
 [\hat \rho_{\bar q,j}, \hatS_{j,\mu} ] \bar{\hatN}_j T_i^a T_j^a  \hatN_j
 \right ),
 \label{eq2.21}
 \end{split} 
\ee
where $\hat \rho_{\bar q,j} = \hatp_j - m_j$.   Since
\be
[\hat \rho_j,\hatS_{j}^\mu ] = L_j^\mu  \hat \rho_j,
\ee
regardless of whether the density matrix refers to a quark or an anti-quark, we observe that eqs.~(\ref{eq2.20}-\ref{eq2.21}) combine with the last term of eq.~(\ref{eq2.17}) into
\be
\sum \limits_{i,j \in \Nset} \eta_i \eta_j J_i^\mu L_{j,\mu} F^{ij}_{\rm LO}.
\ee
We emphasise that the differential operator $L_j^\mu$ in the above equation \emph{does} act  on all $p_j$-dependent terms in $F_{\rm LO}^{ij}$ without any restrictions.

We conclude that the amplitude squared that describes the  emission of soft gluons in the process
$\varnothing \to \sum  \limits_{i \in \Nset} f_i(p_i) + X(p_X)$ with subleading accuracy in $k$ can be written as follows 
\be
|{\cal A}_{\rm real}|^2 = -g_s^2 \sum \limits_{i,j \in \Nset} \eta_i \eta_j W_i^\mu W_{j, \mu} F_{\rm LO}^{ij} + {\cal O}(k^0).
\label{eq2.25}
\ee
In eq.~(\ref{eq2.25}) we introduced the generalised current $W_i^\mu$ which reads
\be
W_i^\mu = J_i^\mu + \frac{1}{2} L_i^\mu,
\label{eq2.26}
\ee
and it is understood that the differential operator $L_i^\mu$ does not act on the eikonal currents $J_i^\mu$ but only on the colour-correlated matrix element $F_{\rm LO}^{ij}$.

Finally, we note that, for processes with some particles in the initial state, $|{\cal A}_{\rm real}|^2$ can be obtained from our result by crossing.  To this end, to describe an initial-state (anti)-particle $i$, one starts with eq.~(\ref{eq2.25}) and inverts the corresponding momentum $p_i \to -p_i$ in the definitions of  $J_i$, $D_i$, $L_i$ and $d_i$. In addition, one needs to set $\eta_i=-1$ if $i$ is an initial-state quark and $\eta_i=1$ if $i$ is an initial-state anti-quark. This completes the discussion of the real-emission part.

\subsection{Virtual corrections}
\label{sect:virt}

\begin{figure}
    \centering
    \begin{subfigure}[t]{0.32\textwidth}
        \includegraphics[width=1\textwidth]{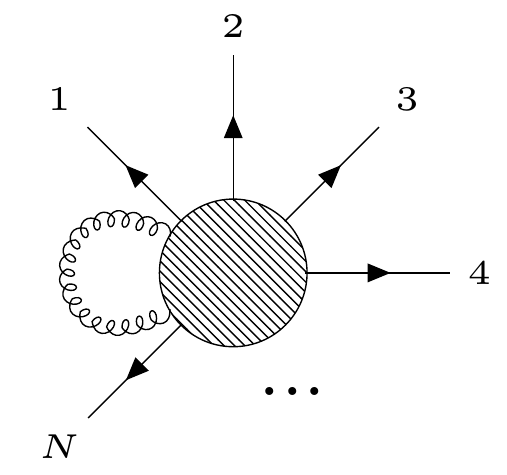}
    \end{subfigure}
    \hfill
    \begin{subfigure}[t]{0.32\textwidth}
        \includegraphics[width=1\textwidth]{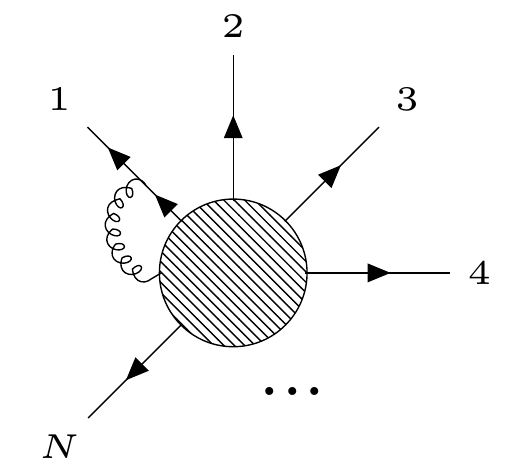}
    \end{subfigure}
    \hfill
    \begin{subfigure}[t]{0.32\textwidth}
        \includegraphics[width=1\textwidth]{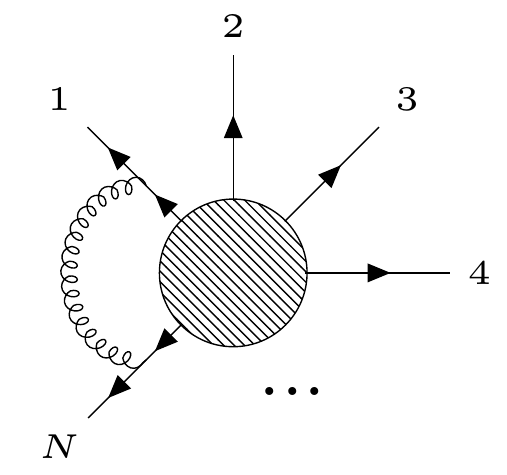}
    \end{subfigure}
    \caption{Loop contributions that need to be considered. The first diagram belongs to group $V_1$, while the second and third diagrams belong to groups $V_2$ and $V_3$ respectively (see text for details).}
    \label{fig2}
\end{figure}

We need to analyse the virtual corrections in a similar way. The one-loop diagrams that contribute to the process of eq.~(\ref{eq1.1a}) can be divided into three distinct groups. The first group includes diagrams where the virtual gluon is not connected to any of the external lines.
The second group comprises all diagrams where the virtual gluon is attached to one and only one external line.
The third group includes diagrams where the virtual gluon connects two external on-shell lines. These different contributions are shown in Fig.~\ref{fig2}. We will analyse each one of them in turn.
We note that we will also have to include the wave function renormalisation contribution that corresponds to self-energy insertions on the external lines and account for mass counterterms. 

It is straightforward to convince oneself that diagrams of the first group cannot contain ${\cal O}(\lambda)$ terms. However, this is not the case for the diagrams in the second and third groups. To analyse the diagrams that belong to the second group, we note that their sum can be written in the following way
\be
   | {\cal A}_{V_2} \rangle =
   \int \frac{{\rm d}^4 k}{(2 \pi)^4} \frac{-i}{k^2 - \lambda^2} \; | {\cal M}_{V_2} \rangle,
   \ee
where 
\begin{equation}
    \begin{split}
      | {\cal M}_{V_2} \rangle
      =& g_s^2 \sum_{i \in \qset } \bar{u}_i \gamma_\mu \frac{\hatp_i + \hatk + m_i}{d_i} T^a_i \hatN^{a,\mu}_i(p_1,p_2,..,p_i + k,...|-k) \\
       +& g_s^2 \sum_{i \in \qbarset }   \hatN^{a,\mu}_i(p_1,p_2,..,p_i + k,...|-k) T^a_i \frac{-\hatp_i - \hatk + m_i}{d_i}  \gamma_\mu v_i.
    \end{split}
    \end{equation}
    In the above equation  $ d_i = (p_i + k)^2 - m_i^2$  
    and $\hatN^{a,\mu}_i(p_1,p_2,..,p_i + k,...|-k)$ is the Green's function that describes the structure-dependent radiation in the process of eq.~(\ref{eq1.1b}) with the off-shell leg $i$. A simple power counting shows that, for the purpose of computing ${\cal O}(\lambda)$ contributions to the virtual amplitude, the above expression can be simplified to
    \be
    \begin{split}
      | {\cal M}_{V_2} \rangle   &=  g_s^2 \sum_{i \in \qset}  J_{i,\mu}   \; T^a_i \bar{u}_i \hatN^{a,\mu}_i(p_1,p_2,..,p_i,...|0)
      \\
       &  - g_s^2 \sum_{i \in \qbarset }   J_{i,\mu} \hatN^{a,\mu}_i(p_1,p_2,..,p_i,...|0) \; v_i  T^a_i 
        + {\cal O}(k^0), 
    \end{split}
\end{equation}
where $J_{i}^\mu = 2 p_i^\mu/d_i$ is the eikonal current and it is indicated that we need the function $\hatN_i^{a,\mu}$ at $k = 0$. To compute it, we proceed similarly to what was done in the previous section and write
\be
g_s \hatN^{a,\mu}_{i_{\bar q}}(p_1,p_2,..,p_i,...|0) \; v_{i_{\bar q}} = g_s \bar{u}_{i_q}  \hatN^{a,\mu}_{i_q}(p_1,p_2,..,p_i,...|0)
= {\cal M}^{a,\mu}_{\rm reg}(p_1,p_2,..,p_N|0),
\ee
for all $i_q$ and $i_{\bar q}$, and with  ${\cal M}^{a,\mu}_{\rm reg}$ given in  eq.~(\ref{eq2.12}). Using eq.~(\ref{eq2.12}), we find
\begin{equation}
    \begin{split}
     | {\cal M}_{V_2} \rangle  = -g_s^2 \sum_{i ,j \in \Nset} \eta_i \eta_j \; J^\mu_i {\bar D}_{j,\mu} \;  T^a_i T^a_j \; | {\cal M}_0 \rangle .
      \label{eq2.30}
    \end{split}
\end{equation}

We will need to compute the interference of the one-loop amplitude with the leading order amplitude.  Using eq.~(\ref{eq2.30}), we find
\be
  \langle {\cal M}_0 | {\cal A}_{V_2} \rangle
 + \langle {\cal A}_{V_2} | {\cal M}_{0} \rangle 
     = 
- g_s^2 
  \int \frac{{\rm d}^4 k}{(2 \pi)^4} \; \frac{-i}{k^2 - \lambda^2} \; \sum_{i ,j \in \Nset} \; \eta_i \eta_j \; J^\mu_i {\bar D}_{j,\mu}  F_{\rm LO}^{ij}.
\ee

Next, we consider gluon exchanges between two external lines. There are three similar but distinct cases that need to be studied, namely the exchanges between two quarks, two anti-quarks and a quark and an anti-quark. We write
\be
| {\cal A}_{V_3} \rangle  =
   \int \frac{{\rm d}^4 k}{(2 \pi)^4} \frac{-i}{k^2 - \lambda^2} \; | {\cal M}_{V_3} \rangle ,
\ee
where 
\begin{equation}
    \begin{split}
      | {\cal M}_{V_3} \rangle
      =& g_s^2\sum_{i_{q}<j_{q}} \bar{u}_i \left( J^\mu_i + \hatS^\mu_i \right) \bar{u}_j
      \left( I_{j,\mu} - \hatC_{j,\mu}  \right) T^a_i T^a_j \hatN_{i j}(...,p_i + k,.., p_j - k,...) 
    \\
    +& g_s^2\sum_{i_{q},j_{\bar{q}}} \bar{u}_i \left(J^\mu_i + \hatS^\mu_i \right)
    T^a_i T^a_j \hatN_{i j}(...,p_i + k,.., p_j - k,...) \left(-I_{j,\mu} - \hatC_{j,\mu}  \right)  v_j
        \\
    +& g_s^2\sum_{i_{\bar{q}}<j_{\bar{q}}} T^a_i T^a_j \hatN_{i j}(...,p_i + k,..,p_j - k,...)
    \left(-I_{j,\mu} - \hatC_{j,\mu}  \right)  v_j \left(-J^\mu_i + \hatS^\mu_i \right)  v_i .
    \end{split}
\end{equation}
Analogous to the definition of $\hatN_i$, the quantity $\hatN_{i j}(...,p_i + k,.., p_j - k,...)$ is the Green's function of the process in eq.~(\ref{eq1.1a}) with amputated off-shell legs $i$ and $j$.
In the above equation,  $J^\mu_i $ and  $ \hatS^\mu_i$    have already been  defined and  
\be
I_j^\mu  = \frac{2 p^\mu_j - k^\mu}{d^-_j},\;\;\;\;\;\hatC_j^\mu = \frac{\sigma^{\mu \nu} k_\nu}{d^-_j},
\ee
with   $d^-_j = (p_j - k)^2 - m_j^2$.
We expand $| {\cal M}_{V_3} \rangle$ through relevant order in $k$ and find 
\begin{equation}
    \begin{split}
     | {\cal M}_{V_3} \rangle 
      &= \frac{g_s^2}{2} \sum_{i \ne j \in \Nset}  \eta_i \eta_j \left (
      J^\mu_i I_{j,\mu}      +  J^\mu_i I_{j,\mu} \;  k_\nu \;  {\bar \Delta}^\nu_{i j}
      \right )  T_i^a T_j^a  | {\cal M}_0 \rangle \\
        & +  \frac{g_s^2}{2} \sum_{i \ne j \in \Nset}  T^a_i T^a_j\left( \eta_j I_{j,\mu} |{\cal M}_{0,i}^{s,\mu} \rangle 
  -  \eta_i J^\mu_i  | {\cal M}_{0,j}^{\sigma,\mu} \rangle  \right),
    \end{split}
\end{equation}
where ${\bar \Delta}^\nu_{i j} = {\bar D}^{ \nu}_i - {\bar D}^{\nu}_j$ and we used the fact that $I_{i}^\mu(k) = J_{i}^{\mu}(-k)$
and $\hatC_j^\mu(-k) = - \hatS_j^\mu(k)$. $| {\cal M}_{0,j}^{\sigma,\mu} \rangle$ is defined similarly to eq.~(\ref{eq:strucdependM0}) with $\hatS_i^{\mu}$ being replaced by $\hatC_i^{\mu}$.
Next, we need to compute
\be
\langle {\cal M}_0 | {\cal M}_{V_3} \rangle + \langle {\cal M}_{V_3} | {\cal M}_0 \rangle.
\ee
The computation is very similar to what has been done for the real-emission contribution. We recall that the key point is to rewrite the commutators of the spin operators with the quark and anti-quark density matrices as the derivatives of the density matrices with respect to the external momenta, and then combine these derivatives with ${\bar \Delta}_{ij}^\nu$ acting on $F_{\rm LO}^{ij}$. We find 
\begin{equation}
  \begin{split}
    &  \langle {\cal M}_0 | {\cal M}_{V_3} \rangle + \langle {\cal M}_{V_3} | {\cal M}_0 \rangle  
    \\
    & =  \frac{g_s^2}{2}  \sum_{i \ne  j \in \Nset} \eta_i \eta_j \left[ 2 J^\mu_i  I_{j,\mu}  + J^\mu_i  I_{j,\mu} k_\nu {\bar \Delta}^\nu_{i j} + I^\mu_{ j} {\tilde L}_{i,\mu}
      -  J^\mu_i {\tilde K}_{j,\mu} \right]F^{i j}_{\text{LO}},
    \end{split}
\label{eq2.41}
\end{equation}
where
\be
L_i^\mu = J_i^\mu k^\nu D_{i,\nu}-D_{i}^\mu,\;\;\;\; K^\mu_i = I_i^\mu k^\nu D_{i,\nu} + D_{i}^\mu,
\ee
so that
\be
   [\hat{\rho}_i , \hatS^\mu_i] = L^\mu_i \hat{\rho}_i,\;\;\;\;[\hat{\rho}_j,  \hatC_{j,\mu} ] = K_{j,\mu} \hat{\rho}_j, 
   \ee
   and tilde indicates that these differential operators \emph{only} act on the density matrices in $F_{\rm LO}^{ij}$.

   To obtain the final result, we combine eq.~(\ref{eq2.41}) and eq.~(\ref{eq2.30}), where in the latter equation we separate the contributions with $i=j$ from those with $i \ne j$. We find
   \be
   \begin{split}
    &  \langle {\cal M}_0 | {\cal M}_{V_2} + {\cal M}_{V_3} \rangle + \langle {\cal M}_{V_2} +{\cal M}_{V_3}
       | {\cal M}_0 \rangle  
     \\
 &     =
\frac{g_s^2}{2}  \sum_{i \ne  j \in N} \eta_i \eta_j \left[ 2 J^\mu_i  I_{j,\mu}  + J^\mu_i  I_{j,\mu} k_\nu {\bar \Delta}^\nu_{i j} + I^\mu_{ j} {\tilde L}_{i,\mu}
      -  J^\mu_i {\tilde K}_{j,\mu} \right]F^{i j}_{\text{LO}}
\\
& - g_s^2 
\sum_{i \ne j \in N} \eta_i \eta_j\, J^\mu_i {\bar D}_{j,\mu}  F_{\rm LO}^{ij}  - g_s^2 C_F \sum \limits_{i=1}^{N} J^\mu_i {\bar D}_{i,\mu}  F_{\rm LO},
\label{eq2.44}
   \end{split}
   \ee
where in the last term we used $\sum_{a} T_i^a T_i^a = C_F$ for $i \in N$. 

It is convenient to rewrite certain  terms that appear in the above expression. First we note that 
   \begin{equation}
\begin{split}
    J^\mu_i  I_{j,\mu} k_\nu \bar \Delta_{ij}^\nu =& I_{j,\mu} {\bar L}^\mu_i + I_{j,\mu} {\bar D}^\mu_i - J^\mu_i {\bar K}_{j,\mu} +  J^\mu_i {\bar D}_{j,\mu}.
\end{split}
\end{equation} 
   Since ${\bar K}_{i,\mu} + {\tilde K}_{i,\mu} = K_{i,\mu}$ and ${\bar L}_{i,\mu} + {\tilde L}_{i,\mu} = L_{i,\mu}$,
   we find
   \be
J^\mu_i  I_{j,\mu} k_\nu {\bar \Delta}^\nu_{i j} + I^\mu_{ j} {\tilde L}_{i,\mu}
-  J^\mu_i {\tilde K}_{j,\mu}
= I_{j,\mu} L^\mu_i  - J^\mu_i K_{j,\mu} + I_{j,\mu} {\bar D}^\mu_i +  J^\mu_i {\bar D}_{j,\mu}.
   \ee
   Next, we combine the last two terms from the above equation with the next-to-last term in eq.~(\ref{eq2.44}).
   We find
   \be
   \begin{split} 
     &  \frac{g_s^2}{2} \sum \eta_i \eta_j   \left ( I_{j,\mu} {\bar D}^\mu_i +  J^\mu_i {\bar D}_{j,\mu}   - 2 J_i^\mu \bar D_{j,\mu} \right ) F_{\rm LO}^{ij}
     \\
        & = \frac{g_s^2}{2} \sum \eta_i \eta_j   \left ( I_{j,\mu} {\bar D}^\mu_i -  J^\mu_i {\bar D}_{j,\mu}   \right ) F_{\rm LO}^{ij}
   \to 0.
   \label{eq2.47}
  \end{split} 
   \ee
   The last step  follows from the fact that, to compute the one-loop  amplitude, 
   we will have to integrate eq.~(\ref{eq2.47}) over $k$ with the weight $1/(k^2 - \lambda^2)$,  and from the observation that through leading order in $k$,  $I_j^\mu(-k) = J_j^\mu(k)$. Hence, we obtain
\begin{equation}
  \begin{split}
   &  \langle {\cal M}_0 | {\cal M}_{V_2} + {\cal M}_{V_3} \rangle + \langle {\cal M}_{V_2} +{\cal M}_{V_3}
       | {\cal M}_0 \rangle  
     \\
 &     =
\frac{g_s^2}{2}  \sum_{i \ne  j \in N} \eta_i \eta_j \left[ 2 J^\mu_i  I_{j,\mu}  + I_{j,\mu} L^\mu_i  - J^\mu_i K_{j,\mu} \right]F^{i j}_{\text{LO}}
\\
& - g_s^2 C_F \sum \limits_{i=1}^{N} J^\mu_i {\bar D}_{i,\mu}  F_{\rm LO}.
    \end{split}
\end{equation}

It is further convenient to rewrite the last term as follows
\be
\sum_{i} J^\mu_i {\bar D}_{i,\mu} F_{\text{LO}}
 = \sum_{i} J^\mu_i \left ( D_{i,\mu} - {\tilde D}_{i,\mu} \right ) F_{\text{LO}}.
 \ee
 Then
 \be
 \sum_{i} J^\mu_i {\tilde D}_{i,\mu} \;  F_{\text{LO}}
 = \sum_{i} \frac {2 }{d_i} F_{\text{LO}}|_{\hat{\rho}_i = \hatp_i}
 =  \sum_{i} \frac{2}{d_i} F_{\text{LO}} -  \sum_{i} \frac{2 \eta_i}{d_i} F_{\text{LO}}|_{\hat{\rho}_i = m_i \hatone}.
 \ee
 In this equation, the subscripts indicate that the density matrix of a fermion $i$ should be replaced either with $\hatp_i$ or with $m_i$ times the identity matrix \hatone.

Putting everything together and using $I^\mu_j(k) = J^\mu_j(-k)$ and $K^\mu_j(k) = - L^\mu_j(-k)$,  we obtain 
\begin{equation}
    \begin{split}
 \langle {\cal M}_0 | {\cal M}_{V_2} + {\cal M}_{V_3} \rangle + \langle {\cal M}_{V_2} +{\cal M}_{V_3}
      | {\cal M}_0 \rangle  &= \frac{g_s^2}{2}  \sum_{i \ne j \in N} 2 \eta_i \eta_j  W^\mu_i(k) W_{j,\mu}(-k)  F^{ij}_{\text{LO}}
      \\
   & \hspace{-3cm}   -g^2_s C_F  \sum_{i=1}^{N} \left( J^\mu_i D_{i,\mu} F_{\text{LO}} - \frac{2}{d_i} F_{\text{LO}}
      + \frac{2 \eta_i}{d_i} F_{\text{LO}}|_{\hat{\rho}_i = m_i \hatone} \right),
    \end{split}
\end{equation}
where $W^\mu(k)$ is defined in eq.~(\ref{eq2.26}).   The final result for the ${\cal O}(\lambda)$ contribution that originates from
the one-loop amplitude  reads
\be
\begin{split} 
|{\cal M}_V|^2 = & g_s^2  {\cal T}_\lambda \Bigg [ \int \frac{ {\rm d}^4 k}{(2\pi)^4}  \frac{-i}{k^2 - \lambda^2}
\Bigg \{ \sum \limits_{i \ne j \in N}  \; \eta_i \eta_j \;W_i^\mu(k) W_{j,\mu}(-k)  F_{\rm LO}^{ij}
\\
& - C_F \sum \limits_{i=1}^{N} \left (
 J_i^\mu D_{i,\mu} F_{\rm LO} - \frac{2}{d_i} F_{\rm LO} + \frac{2 \bar{\eta}_i }{d_i} F_{\rm LO}|_{\hat \rho_i \to m_i \hatone} \right )
 \Bigg  \}
 \Bigg ],
\end{split} 
\label{eqv2}
\ee
where ${\cal T}_\lambda$ is an operator that extracts the contribution linear in $\lambda$ from the expression it is applied to. We have introduced the quantity $\bar{\eta}_i$ in the above equation to enable crossing to the initial state.  We define  $\bar{\eta}_i$ to be equal to $\eta_i\; (-\eta_i)$ if $i$ is in the final (initial) state. Furthermore, for dipoles involving initial-state particles, we can use the same expression, eq.~(\ref{eqv2}), and we need to apply the same changes as in the real-emission part. This means that the corresponding momenta should be inverted $p_i \to -p_i$ in the definitions of $J_i$, $D_i$, $L_i$ and $d_i$ and the correct $\eta_i$-values for initial-state quarks and anti-quarks have to be assigned.

\section{Connection   to the large-$N_f$ limit of QCD}
\label{sect:3}

There is an important difference between the calculation that we just described and traditional applications of the renormalon calculus. This difference is related to the fact that, in the current case, virtual gluons appear already in the tree-level diagrams.
This leads to the appearance of perturbative corrections that scale as ${\cal O}(\alpha_s(Q) N_f)$ where $Q$ is the hard scale of the process, and $N_f$ is the
number of massless fermions. These corrections are peculiar because the relation between renormalon calculus and calculations where the gluon is assigned a small but non-vanishing mass is derived by considering the $N_f \to -\infty$ limit; in this limit the behaviour of the ${\cal O}(\alpha_s(Q) N_f)$ corrections needs to be clarified. 

\begin{figure}[t]
    \centering
    \begin{subfigure}[t]{.49\linewidth}
        \centering
        \vskip 0pt
        \includegraphics[width=0.7\textwidth]{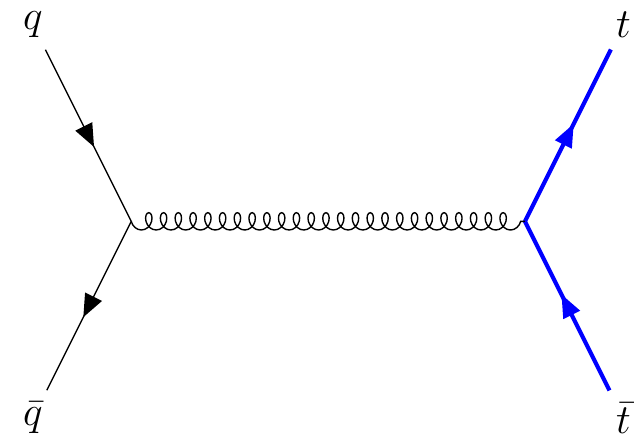}
        \caption{}
    \end{subfigure}\hfill
    \begin{subfigure}[t]{.49\linewidth}
        \centering
        \vskip 0pt
        \includegraphics[width=0.7\textwidth]{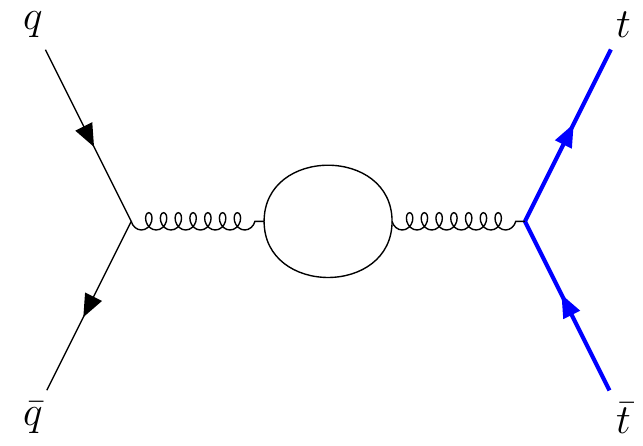}
        \caption{}
    \end{subfigure}\hfill
    \caption{(a) Leading order diagram and (b) $N_f$-dependent vacuum polarisation contribution to  $q \bar q \to t \bar t$ process.}
        \label{fig5}
\end{figure}

To explain the origin of these  corrections, we focus on an example where a top quark pair is produced in the collision of a massless
quark and an anti-quark. At leading order, there is just one diagram that contributes to this process, it is shown in Fig.~\ref{fig5}a.
The one-loop corrections include the vacuum polarisation diagram shown in Fig.~\ref{fig5}b which, together with ${\cal O}(N_f)$ contribution to the strong coupling renormalisation constant,  combine to fix the scale of $\alpha_s$ 
in the leading order amplitude to the hard
scale $Q^2 = (p_q + p_{\bar q})^2$.  Moreover, if we choose the  
so-called $V$-scheme to renormalise the strong coupling constant, we can absorb the entire $N_f$-dependent vacuum polarisation
contribution into the strong coupling constant \cite{Peter:1996ig, Schroder:1998vy}. The leading order matrix element then reads 
\be
   {|\cal M}_0 \rangle = \frac{ i 4 \pi  \alpha_{s,V}(Q)  }{Q^2} \left [ \bar v (p_{\bar q})  \gamma^\mu T^a u(p_q)
     \right ]\; \left [ \bar u_t(p_t) T^a \gamma^\mu v_{\bar{t}}(p_{\bar{t}}) \right ].
\label{eq3.1}
\ee

\begin{figure}
    \centering
    \begin{subfigure}[t]{.49\linewidth}
        \centering
        \vskip 0pt
        \includegraphics[width=0.7\textwidth]{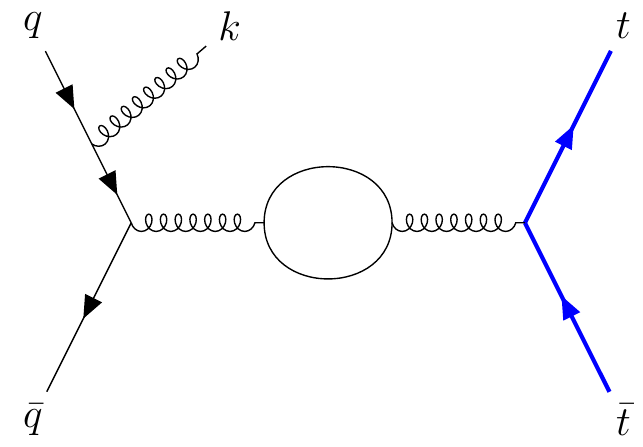}
    \end{subfigure}\hfill
    \begin{subfigure}[t]{.49\linewidth}
        \centering
        \vskip 0pt
        \includegraphics[width=0.7\textwidth]{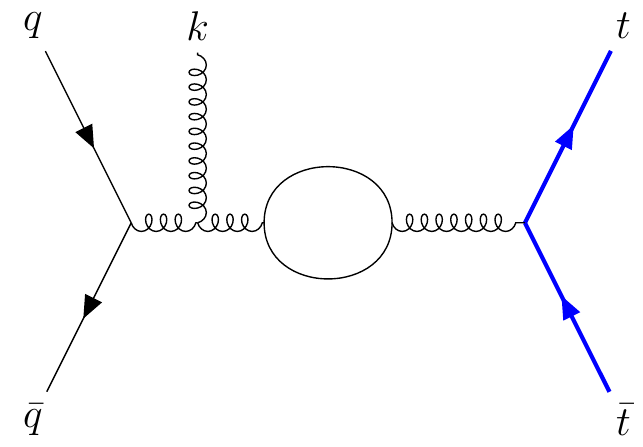}
    \end{subfigure}\hfill

    \begin{subfigure}[t]{.49\linewidth}
        \centering
        \vskip -15pt
        \includegraphics[width=0.7\textwidth]{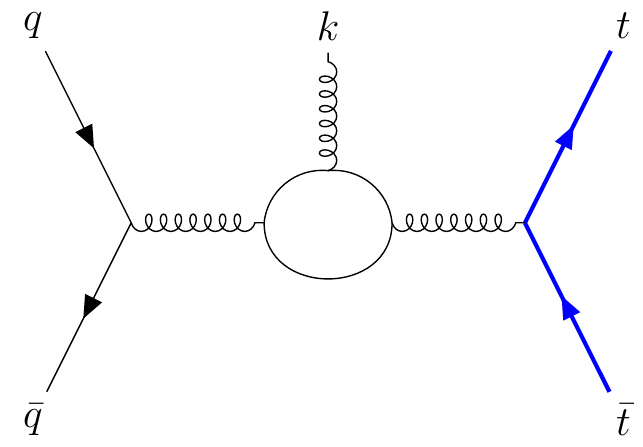}
    \end{subfigure}\hfill
        \caption{Examples of  contributions to $q \bar q \to t \bar t$ which are proportional to   
        ${\cal O}(g_s\alpha_s(Q) N_f)$.}
        \label{fig6}
\end{figure}

We now discuss what happens at NNLO and, as an example,  we consider an emission of a soft gluon in  diagrams with the vacuum polarisation insertions. These diagrams are shown in Fig.~\ref{fig6}. The important feature of all these diagrams is that they are \emph{hard}, in the sense that the $k \to 0$ limit does not induce additional singularities in the $N_f$-dependent parts of the diagrams. Therefore, these diagrams can then be studied following the proof of the LBK theorem.

It is then  straightforward to show that all diagrams similar to the ones displayed in Fig.~\ref{fig6} can indeed be obtained from the LBK theorem provided that $|{\cal M}_0 \rangle$ is chosen as in eq.~(\ref{eq3.1}).  Interestingly, this implies that when the structure-dependent radiation is computed by differentiating
the tree-level amplitude as in eq.~(\ref{eq2.12}), the derivative of the strong coupling constant in eq.~(\ref{eq3.1}) also needs to be calculated. However, beyond that, there seem to be no additional implications for the computation of linear power corrections related to the presence of off-shell gluons in the leading order matrix elements. It is very plausible that this result generalises also to higher orders both within and beyond the large-$N_f$ resummation framework; we will, however, leave an all-orders investigation of this factorisation to future work.

\section{Cancellation of ${\cal O}(\lambda)$ contributions to the total cross section}\label{sec:cancellation}

In the previous section we computed the next-to-leading soft terms in the radiative corrections  to a process involving an arbitrary number of external quarks, anti-quarks and colour-neutral particles,  caused by the production or
exchange of a soft massive gluon. This allows us to calculate the expansion of the cross section in the gluon mass $\lambda$ including ${\cal O}(\lambda)$ terms.  The question that we would like to answer 
in this section is whether such terms are present in the total cross section of the  $q \bar q \to t \bar t $ process.

According to the analysis in the previous section, two types of terms appear in the sum of the real and virtual contributions.
First, there are terms that depend on a particular colour-correlated amplitude 
squared $F_{\rm LO}^{ij}$.
We will refer to such contributions as ``dipole''.  Second,
there are terms which depend on the leading order amplitude squared
$F_{\rm LO}$ multiplied by the Casimir operator $C_F$. We will refer to such terms as ``monopole''. 
We will show that the cancellation of the ${\cal O}(\lambda)$ terms takes place individually for each of the dipole and monopole terms, and therefore it is convenient to study them separately.

In this respect, we note that, for processes with massive quarks, the cancellation of the ${\cal O}(\lambda)$ terms requires us to introduce the renormalisation of the quark mass parameter and the wave function renormalisation. In addition, it is to be expected that the cancellation requires us to express the cross section in terms of a mass parameter that is free of ${\cal O}(\lambda)$ terms (see e.g. ref.~\cite{Makarov:2023ttq} for a related analysis). Thus, if the calculation is performed in the on-shell mass scheme, one has to rewrite the leading order cross section in terms of the new mass parameter. Since all these renormalisation factors and mass shifts are  proportional to the Casimir factor $C_F$, these contributions will have to be added to the monopole terms to ensure the cancellation.

In what follows, we will explicitly study the $q \bar q \to t \bar t + X$  process. There are six dipoles ($q \bar q$, $q t$, $ q \bar t$, $\bar q t$, $\bar q \bar t$ and $t \bar t$), and four monopoles ($qq$, $\bar q \bar q$, $tt$ and $\bar t \bar t$) to  consider.
Since $q$ and $\bar q$ are massless, according to ref.~\cite{Caola:2021kzt},
the corresponding dipole $q \bar q$ and the two monopoles $qq$ and $\bar q \bar q$ do not produce ${\cal O}(\lambda)$ terms and can be discarded. We will also make use of the fact that masses of $t$ and $\bar t$ are identical which allows us to combine the $tt$ and $ \bar t \bar t$ monopoles into a single contribution. Hence, we write 
\be
\begin{split} 
   {\cal T}_\lambda \left [ {\rm d} \sigma(q \bar q \to t \bar t+X) \right ]
   & =     {\cal T}_\lambda \left [  {\rm d} \sigma_R^{ t \bar t } + {\rm d} \sigma_V^{ t \bar t} \right ] +
   \sum \limits_{f_1 = q,\bar q} \sum \limits_{f_2 = t,\bar t}
    {\cal T}_\lambda \left [  {\rm d} \sigma_R^{ f_1 f_2 } + {\rm d} \sigma_V^{ f_1 f_2} \right ]
         \\
  &  + \sum \limits_{f = t,\bar t} {\cal T}_\lambda \left [  {\rm d} \sigma^{ff}_{R} + {\rm d}\sigma^{ff}_V + {\rm d}\sigma_{\rm ren} + {\rm d}\sigma_{\rm mass} \right ],
   \end{split} 
\ee
where in the first line we collected the various dipole contributions and in the second line the two monopole contributions together with terms generated by the renormalisation and mass redefinition. We will now proceed with the analysis of the various terms in the above equation. However, before we dive into this discussion, we will have to describe the momenta mappings required to enable the integration over the gluon momentum in the real-emission contributions. 

\subsection{Momenta mappings}
\label{sec:momentummapping}

We consider the process $q(p_q) + \bar q(p_{\bar q}) \to t(q_t) + \bar t( q_{\bar t}) + X(p_X) + g(k)$. In order to integrate out the gluon momentum $k$ and express the result in terms of the LO cross section in a process-independent manner, we will need to factorise out the gluon momentum. This will allow us to combine real emission and virtual contributions in a convenient manner. To remove the momentum
of the gluon from the delta-function  that enforces the energy-momentum conservation, we change the momenta of top quark
and anti-quark. Specifically, we write
\begin{equation}
    \begin{split}
        q_t =& \; p_t - \alpha k + A(k) p_t + B(k) p_{\bar t}, \\
        q_{\bar t} =& \; p_{\bar t} - \beta k - A(k) p_t  - B(k) p_{\bar t},
    \end{split}
    \label{eq:momentummapping}
\end{equation}
where two parameters $\alpha$ and $\beta$ are $k$-independent, and $A$ and $B$ are two ${\cal O}(k)$ functions. 
The mapping must satisfy the condition
\be
q_t + q_{\bar t}  + k  = p_t + p_{\bar t},
\ee
which implies
\be
1  = \alpha + \beta.
\label{eqalphabeta}
\ee
Furthermore, imposing the conditions
\be
p_t^2 = p_{\bar t}^2 = q_t^2 = q_{\bar t}^2 = m_t^2,
\ee
we find the following results 
\be
\begin{split} 
A & =  -\frac{\alpha \,  m_t^2 \, (p_t k)  + \beta \, (p_t p_{\bar t}) \, (p_{\bar t} k)}{(p_t p_{\bar t})^2 - m_t^4},\\
B & = \frac{\alpha \,  (p_t p_{\bar t})  \, (p_t k)  + \beta \, m_t^2 \, (p_{\bar t} k)}{ (p_t p_{\bar t})^2 - m_t^4}.
\end{split} 
\label{eq:mappingconstraint}
\ee

Using the mapping in eq.~(\ref{eq:momentummapping}), it is straightforward to find the phase space transformation. Keeping ${\cal O}(k)$ contributions
and neglecting higher order terms, we obtain 
\be
\begin{split} 
   & {\rm dLips}(p_q,p_{\bar q}; q_t,q_{\bar t},p_X, k)
    = {\rm dLips_{LO}}(p_q,p_{\bar q}; p_t,p_{\bar t},p_X) \frac{{\rm d}^4 k}{(2 \pi)^4} \delta(k^2 - \lambda^2) \times 
   \\
   & \hspace{0.5cm} \times \Bigg  ( 1 +   \frac{  \left[(p_t p_{\bar t}) \left( (p_{\bar t} k) - (p_t k) \right) \left( \alpha - \beta \right)
     - 2 m_t^2 \left( \alpha \, (p_t k)  + \beta \, (p_{\bar t} k)   \right )\right]}{(p_t p_{\bar t})^2 - m_t^4}
    \Bigg ).
\end{split} 
\ee

The momenta mappings shown in eq.~(\ref{eq:momentummapping}) will have to be applied to the leading order matrix element squared $F_{\rm LO}$ that appears in both the dipole and monopole terms. Working through the first order in $k$, we find
\be
\label{eq:Ftrans}
F_{\rm LO}(q_t,q_{\bar t}) = \Big[
    1 + (A p_t^\nu + B p_{\bar t}^\nu - \alpha k^\nu) D_{t,\nu}
    -  (A p_t^\nu + B p_{\bar t}^\nu  + \beta  k^\nu) D_{\bar{t},\nu} 
\Big]F_{\rm LO}(p_t,p_{\bar t}).
\ee

The mappings shown in eq.~\eqref{eq:momentummapping} also affect the $t$ and $\bar t$ propagators that appear explicitly in the eikonal currents. The expansion of these propagators through linear terms in $k$ is straightforward and we do not present it here.

\subsection{Individual dipole and monopole contributions to the $t \bar t$ cross section}
\label{sect:final}

In the following, we will discuss the individual dipole and monopole contributions to the $q \bar q \to t \bar t + X$ processes.

\subsubsection{The case of the $t \bar{t}$ dipole}

We start by considering the contribution of the $t \bar t$ dipole to the cross section. In this case, we only need
to combine the real-emission contribution and the contribution of  the virtual corrections.
As explained earlier, no renormalisation contributions  need to be added in this case.

We use eq.~(\ref{eq2.25}) to write the real-emission contribution in the following way
\begin{equation}
\begin{split}
  {\cal T}_\lambda \left[ {\rm d} \sigma_{R}^{t\bar t} \right ]
   &= {\cal T}_\lambda \bigg[ g_s^2 \int {\rm d Lips}(p_q,p_{\bar q}; q_t,q_{\bar t},p_X,k) \times
   \\
   & \times \left (
   2 J^\mu_{t} J_{ \bar t, \mu } + J^\mu_{t} L_{\bar t, \mu} + J^\mu_{\bar t} L_{t, \mu} \right) F_{\rm LO}^{t \bar t}(q_t,q_{\bar t}) \bigg].
\end{split}
\end{equation}
We then perform the momentum transformation using the formulas in the previous section, integrate over $k$ with the help of the phase-space integrals collected in Appendix~\ref{app:phaseintegrals}, and find 
\begin{equation}
  \begin{split}\label{eq:TlRttbar}
    {\cal T}_\lambda & \left [ {\rm d} \sigma_{R}^{t\bar t} \right ]  =
    -\frac{\alpha_s}{2 \pi} \frac{\pi \lambda}{\mt} \int {\rm d Lips}_{\rm LO}(p_q,p_{\bar q}; p_t,p_{\bar t},p_X) \times  \frac{1}{(p_t p_{\bar t} - m_t^2)} \times  
    \\
    & \times \Big [ 2 ( m_t^2 - 2 p_t p_{\bar t}) +\mt^2(p_{\bar{t},\nu} D_t^\nu+p_{t,\nu} D_{\bar t}^\nu)-(p_t p_{\bar t})
       ( p_{t,\nu} D_t^\nu + p_{\bar{t},\nu} D_{\bar t}^\nu) \Big ]
    F^{t \bar t}_{\rm LO}(p_t,p_{\bar t}).
\end{split}
\end{equation}
We note that this result is obtained for arbitrary $\alpha$ and $\beta$, subject to the constraint $\alpha+\beta=1$. We observe that the dependence on these parameters has disappeared from the final result. 

The contribution from the virtual corrections can be extracted from the general formula in eq.~(\ref{eqv2}).
In this case, no momentum mapping is involved and one can integrate the relevant expression over the four-momentum
of the virtual gluon. The relevant integrals are collected in Appendix~\ref{apploopi}. We find 
\begin{equation}
  \begin{split}
        {\cal T}_\lambda \left [ {\rm d} \sigma_{V}^{t\bar t} \right ]  & = 
       {\cal T}_\lambda \bigg [ -\gs^2 \int {\rm d Lips}_{\rm LO}(p_q,p_{\bar q}; p_t,p_{\bar t},p_X)   \int \frac{d^4 k}{(2\pi)^4}
        \frac{-i}{k^2-\lambda^2} \times 
        \\
        & \times \left  ( 2 J_{\bar t}^\mu(k) J_t^\mu(-k) + J_{\bar t}^\mu(k) L_{t,\mu}(-k)
         + J_t^{\mu}(-k) L_{\bar{t},\mu}(k) \right ) F^{t \bar t}_{\rm LO}(p_t, p_{\bar t}) \bigg]
\\
& = -\frac{\alpha_s}{2 \pi} \frac{\pi \lambda}{\mt}  \int {\rm d Lips}_{\rm LO}(p_q,p_{\bar q}; p_t,p_{\bar t},p_X) \times \frac{1}{(p_t p_{\bar t} - m_t^2)} \times 
\\
& \times \bigg[ 2 ( 2p_t p_{\bar t} - m_t^2)-m_t^2(p_{\bar{t},\nu} D_t^\nu+p_{t,\nu} D_{\bar t}^\nu) 
\\
& \hspace{0.6cm}  + (p_t p_{\bar t})
       ( p_{t,\nu} D_t^\nu + p_{\bar{t},\nu} D_{\bar t}^\nu) \bigg ]  F^{t \bar t }_{\rm LO}(p_t, p_{\bar t}).
\end{split}
\end{equation}
Combining  the above  results for the real and virtual corrections, we obtain
\be
   {\cal T}_\lambda \left [ {\rm d} \sigma_{R}^{t\bar t} \right ]
   + 
{\cal T}_\lambda \left [ {\rm d} \sigma_{V}^{t\bar t} \right ] = 0.
\ee

\subsubsection{The case of the $t q$ dipole}

We continue with the discussion of the $tq$ dipole. In principle, the calculation is very similar to the one for the $t \bar t$ dipole but there is a subtlety related to the fact that the momentum mapping, eq.~(\ref{eq:momentummapping}), involves the momentum of $\bar t$, that does not belong to the $tq$ dipole.
The consequence of this is the appearance of the derivative with respect to the $\bar t$ momentum in the real-emission contribution. However, such a derivative does not appear in the virtual correction to the $tq$ dipole because no momentum mapping is required there.
Hence, the minimal requirement for the cancellation of the ${\cal O}(\lambda)$ corrections to occur in the sum of the real and virtual contributions to the $tq$ dipole (independently of other dipoles and monopoles) is the disappearance of the $\partial F_{\rm LO}^{tq}/\partial p_{\bar t}^\mu $ term after the integration over
$k$ in the real-emission contribution. 

To understand how this can be arranged, we consider eq.~\eqref{eq:Ftrans}, which is the only source
of derivatives w.r.t. $p_{\bar t}^\mu$. Since the coefficient of this derivative is already $O(k)$, we
conclude that the potentially offending term reads
\be
\label{eq:beta_contribution}
{\cal T}_\lambda \left[ \int \frac{ {\rm d}^4 k}{(2 \pi)^4} \delta(k^2 - \lambda^2)
J_{t}^\mu J_{q,\mu}  (A p_t^\nu + B p_{\bar t}^\nu  + \beta  k^\nu) D_{\bar{t},\nu} 
\right] F^{t q}_{\rm LO}, 
\ee
where the two eikonal currents should be taken at  leading power. We would like the above expression to vanish after the integration over $k$. To see how this can occur, we note
that $A$ and $B$ are linear combinations of $ \alpha(p_t k)$ and $\beta (p_{\bar t} k)$.  Since
\be
{\cal T}_\lambda  \left[ \int \frac{ {\rm d}^4 k}{(2 \pi)^4} \delta(k^2 - \lambda^2) J_{t}^\mu J_{q,\mu} \;  (p_t k) \right] = 0,
\ee
we conclude that the non-vanishing contribution in eq.~(\ref{eq:beta_contribution}) is proportional to $\beta$
\be
{\cal T}_\lambda \left[ \int \frac{ {\rm d}^4 k}{(2 \pi)^4} \delta(k^2 - \lambda^2)
J_{t}^\mu J_{q,\mu}  (A p_t^\nu + B p_{\bar t}^\nu  + \beta  k^\nu) D_{\bar{t},\nu} 
F^{t q}_{\rm LO} \right] \sim \beta \; V^\nu \; D_{\bar{t},\nu} \; F^{t q}_{\rm LO}, 
\ee
where $V^\nu$ is a non-vanishing vector  that arises as the result of the integration over $k$. 
Hence, the only way to remove this term from the real emission 
contribution to the $tq$ dipole
is to choose a mapping with $\beta = 0$.  It is important to stress that, although
choosing $\beta = 0$ is a necessary condition, it is not obvious that it  is a sufficient one to ensure a 
cancellation of the ${\cal O}(\lambda)$ corrections within the $tq$
dipole independently of all other contributions. 
However, an explicit calculation shows that this is the case. 

To illustrate this point, we choose $\beta= 0$ and compute the real-emission contribution to the $tq$ dipole. Since the $J_q$ and $L_q$ are defined for the outgoing momenta, we will need to invert the momentum of the initial-state quark in their definitions. In addition, we need to set $\eta_q=-1$.
We then find that
\begin{equation}
\begin{split}\label{eq:TlRtq}
{\cal T}_\lambda \left [ {\rm d} \sigma_{R}^{tq } \right ]  & = 
{\cal T}_\lambda \bigg [ \gs^2 \int {\rm d Lips}(p_q,p_{\bar q}; q_t,q_{\bar t},p_X,k)
\frac{d^4 k}{(2\pi)^3} \delta_{+}(k^2 - \lambda^2) 
\\
& \times \left( 2 J^\mu_{t} J_{q,\mu} + J^\mu_{t} L_{q, \mu} + J^\mu_{q} L_{t, \mu} \right) F^{t q}_{\rm LO}(q_t,q_{\bar t}) \bigg]
\\
& = \frac{\alpha_s}{2 \pi} \frac{\pi \lambda}{\mt} \int {\rm d Lips}_{\rm LO}(p_q,p_{\bar q}; p_t, p_{\bar t},p_X)
\Bigg  (
2 - \frac{m_t^2}{p_t p_q}+ p_{t,\nu} D_t^\nu 
\\
& \;\;\;\;\;\;\; - \frac{\mt^2}{p_t p_q} p_{q,\nu} ( D_q^\nu + D_t^\nu) 
\Bigg  ) F^{t q}_{\rm LO}(p_t,p_{\bar t}).
\end{split}
\end{equation}

A straightforward computation of the virtual corrections, using the integrals presented in Appendix~\ref{apploopi}, gives
\begin{equation}
  \begin{split}
    {\cal T}_\lambda \left [ {\rm d} \sigma_{V}^{tq } \right ] &  = 
    {\cal T}_\lambda \bigg [ -\gs^2 \int {\rm d Lips}_{\rm LO}(p_q,p_{\bar q}; p_t,p_{\bar t},p_X) \frac{d^4 k}{(2\pi)^4} \frac{-i}{k^2-\lambda^2} 
    \\
&    \times \left( 2 J^\mu_{q}(k) J_{t,\mu}(-k) + J_{q,\mu}(k) L_{t,\mu}(-k) + J_{t,\mu}(-k) L_{q,\mu}(k)
    \right )F^{t q}_{\rm LO}(p_t,p_{\bar t}) \bigg] 
    \\
&    = \frac{\alpha_s}{2 \pi} \frac{\pi \lambda}{\mt} \int {\rm d Lips}_{\rm LO}(p_q,p_{\bar q}; p_t,p_{\bar t},p_X)
\Bigg  ( -2 + \frac{m_t^2}{p_t p_q} - p_{t,\nu} D_t^\nu 
\\
&  \;\;\;\;\;\;\; + \frac{\mt^2}{p_t p_q} p_{q,\nu} ( D_q^\nu + D_t^\nu)
\Bigg  )
     F^{t q}_{\rm LO}(p_t,p_{\bar t}).
\end{split}
\end{equation}

Combining the above results, we find
\be
   {\cal T}_\lambda \left [ {\rm d} \sigma_{V}^{tq } \right ] + {\cal T}_\lambda \left [ {\rm d} \sigma_{R}^{tq } \right ]
    = 0.
\ee

\subsubsection{Remaining dipoles}

The remaining dipoles $t \bar q$, $\bar t \bar q$ and $\bar t q$ can be analysed in the same way as the $tq$
dipole. For all of them we use the momentum mapping of eq.~(\ref{eq:momentummapping}).
The cancellations of the ${\cal O}(\lambda)$ terms occur independently for each of these dipoles if we choose $\beta = 0$ for the $\bar q t$ and $\alpha = 0$ for the $q \bar t$ and $\bar q \bar t$ dipoles.
This completes the discussion of the cancellation of ${\cal O}(\lambda)$ terms for all of the dipoles that potentially contribute to the
$q \bar q \to t \bar t+X$ partonic process.

\subsubsection{The monopole $tt$ + $\bar{t}\bar{t}$  contributions}

The last contributions that we need to consider are the monopole contributions, related to the $t$ and $\bar t$ quarks in the final state. In principle, one can design a procedure that deals with each of them separately but, for simplicity, we will consider both of them at once. The main difference with respect to the dipole contributions is the need to account for the renormalisation and to redefine   
the top quark  mass. Hence, the pattern of cancellations becomes more involved. 

The real-emission contribution reads
\begin{equation}
\begin{split}
   {\cal T}_\lambda \left [ {\rm d} \sigma_R^{t t} + {\rm d} \sigma_R^{ \bar t \bar t}
    \right ]
  & =    {\cal T}_\lambda \bigg [ - C_F \gs^2 \;
  \int {\rm d Lips}(p_q,p_{\bar q}; q_t,q_{\bar t},p_X,k)
  \big (J^\mu_{t} J_{t, \mu} + J^\mu_{\bt} J_{\bt,\mu}
\\
& \;\;\;\;\;\;\; + J^\mu_{t} L_{t, \mu }  + J^\mu_{\bt} L_{\bar{t},\mu} \big)  F_{\rm LO}(q_t,q_{\bar t}) \bigg]
\\
&   = \frac{\alpha_s \Cf}{2 \pi} \frac{\pi \lambda}{\mt}
\int {\rm d Lips}_{\rm LO}(p_q,p_{\bar q}; p_t,p_{\bar t},p_X) \times \frac{1}{(p_t p_{\bar t} - m_t^2)} \times 
\\
& \times \bigg  [
  \mt^2 (- 1 + p_{\bar{t},\nu} D_t^\nu + p_{t,\nu} D_{\bar t}^\nu)
\\
& \;\;\;\;\;\;\; -(p_t p_{\bar t})  (1 + p_{t,\nu}  D_t^\nu + p_{\bar{t},\nu} D_{\bar t}^\nu ) \bigg  ]
F_{\rm LO}(p_t,p_{\bar t}),
\end{split}
\label{eqttreal}
\end{equation}
where the dependence on $\alpha$ and $\beta$ has cancelled out.

The virtual corrections evaluate to 
\begin{equation}
  \begin{split}
     & {\cal T}_\lambda \left [ {\rm d} \sigma_V^{t t} + {\rm d} \sigma_V^{ \bar t \bar t}
       \right ] = {\cal T}_\lambda \bigg [ -C_F \gs^2  \; \int {\rm d Lips}_{\rm LO} \frac{d^4 k}{(2\pi)^4} \frac{-i}{k^2-\lambda^2} \times 
       \\
       & \times \bigg( \left( J^\mu_t D_{t, \mu }  + J^\mu_\bt D_{\bt,\mu } - \frac{2}{d_t}  - \frac{2}{d_\bt} \right ) F_{\rm LO} + \frac{2}{d_t} F_{\rm LO}|_{\hat{\rho}_t = m_t \hatone} - \frac{2}{d_\bt} F_{\rm LO }|_{\hat{\rho}_{\bar t} = m_t \hatone} \bigg) \bigg]
\\
&  = \frac{\alpha_s \Cf}{2 \pi} \frac{\pi \lambda}{\mt} \int {\rm d Lips}_{\rm LO} \Big[ (- 2 + p_{t,\nu} D_t^\nu + p_{\bar{t},\nu} D_{\bar t}^\nu)  F_{\rm LO} + F_{\rm LO}|_{\hat{\rho}_t = m_t \hatone} - F_{\rm LO}|_{\hat{\rho}_{\bar t} = m_t \hatone}  \Big].
\end{split}
\label{eqttvirt}
\end{equation}

The above results for the real and  virtual corrections have  to be supplemented with the renormalisation
contributions since they are proportional to the Casimir invariant $C_F$ 
and the leading order 
amplitude squared $F_{\rm LO}$.
The computation is analogous to the single top production case discussed in 
ref.~\cite{Makarov:2023ttq}.  We obtain 
\begin{equation}
  \begin{split} 
    {\cal T}_\lambda \left [ {\rm d} \sigma_{\rm ren}  \right ]
      & = 
    \frac{\alpha_s \Cf}{2 \pi }\frac{\pi \lambda}{\mt}
    \int  {\rm d Lips}_{\rm LO}
     \Bigg  [  3  F_{\rm LO}
     +
    \mt {\rm Tr} \left [ \hat{\rho}_t \frac{ \partial \hatN}{\partial \mt} \hat{\rho}_{\bar{t}} \bar{\hatN} \right ]
   \\
&     +
    \mt {\rm Tr} \left [ \hat{\rho}_t \hatN  \hat{\rho}_{\bar{t}} \frac{ \partial \bar{\hatN}}{\partial \mt}  \right ]
   \Bigg  ],
  \end{split} \label{eq:rencontr}
\end{equation}
where derivatives w.r.t. the mass parameter $m_t$ in the last two terms arise because of the on-shell
counterterm mass insertions on  the internal lines. 

Finally, for the cancellation of the ${\cal O}(\lambda)$ terms, it is 
necessary to express  the cross section through a short-distance mass parameter.
To do this, it  is important to recognise  that the dependence 
of the cross section on the top quark  masses arises in two distinct ways:
1) through the \emph{explicit} appearance of $m_t$ in the matrix elements and 2) through the \emph{implicit} dependence of the momenta of the final state particles on $\mt$.

The explicit dependence is accounted for  by writing  
$\mt = \tilde{m}_t + \delta \mt$ in the function $F_{\rm LO}$ 
and then expanding in $\delta \mt$ to first order.
The corresponding change in the leading order cross section reads 
\begin{equation}
\begin{split} 
  \delta \sigma_{\rm mass}^{\rm expl}
  =& 
  \delta \mt \int  {\rm d Lips}_{\rm LO} \frac{\partial F_{\rm LO}  }{\partial \mt} 
  \\
  =&
  \delta \mt \int  {\rm d Lips}_{\rm LO}  \; \Bigg( 
   {\rm Tr} \left [\hatone \hatN \hat{\rho}_{\bar{t}} \bar{\hatN} \right ]+{\rm Tr} \left [\hat{\rho}_{t}  \hatN (-\hatone) \bar{\hatN} \right ]
  \\ 
   &\hspace{2cm} + {\rm Tr} \left [ \hat{\rho}_{t} \left (
     \frac{ \partial \hatN}{\partial \mt} \hat{\rho}_{\bar{t}} \bar{\hatN} 
           + \hatN \hat{\rho}_{\bar{t}} \frac{ \partial \bar{\hatN}}{\partial \mt}  \right ) \right ]
           \Bigg).
\end{split} 
\end{equation}

The change in the cross section due to the dependence of the momenta of the final-state particles on $\mt$ can be computed by redefining the momenta of the top quark and the anti-top quark as follows 
        \begin{equation}
\ptop = (1-\kappa) \tildeptop + \kappa \tildeptopbar,\;\;\; \ptopbar = (1-\kappa) \tildeptopbar + \kappa \tildeptop,
        \end{equation}
where $\kappa$ is ${\cal O}(\lambda)$. From this, it follows that 
        \begin{equation}
\ptop^2 = \mt^2 = \tildeptop^2 + 2 \kappa  \left(\tildeptop \tildeptopbar - \tildemt^2\right)+{\cal O}(\kappa^2).
        \end{equation}
        Thus, by choosing
        \begin{equation}
\kappa =  \frac{\delta \mt^2}{2\left(\tildeptop \tildeptopbar - \tildemt^2\right)},
\end{equation}
the mass-shell condition for $\tildeptop$ becomes 
\begin{equation}
\tildeptop^2 = \tildemt^2 = \mt^2 - \delta \mt^2. 
\end{equation}

Following the discussion of the momenta mapping  of the real-emission  contribution in Section~\ref{sec:momentummapping} and slightly modifying it where necessary, we obtain 
\begin{equation}
   {\rm dLips}\left ( p_q, p_{\bar q}; \ptop,\ptopbar ,p_X; \mt^2 \right ) =
   {\rm d Lips} \left ( p_q, p_{\bar q}; \tildeptop,\tildeptopbar ,p_X; \tildemt^2 \right ) \left ( 1-2 \kappa +\mathcal{O}{\left(\lambda^2\right)} \right ).
\end{equation}
Finally, expanding the leading order amplitude squared, we determine the change of the cross section due to the implicit mass change 
\begin{equation} \label{eq:implicitmass}
\begin{split} 
& \delta  \sigma^{\rm impl}_{\rm mass} = 
\int {\rm d Lips} \left (..., \tildeptop,\tildeptopbar,.. \right )
\left [-2\kappa  - \kappa \left(\tildeptop^\mu-\tildeptopbar^\mu\right) \left (\frac{\partial }{\partial \tildeptop^\mu}
  - \frac{\partial }{\partial \tildeptopbar^\mu}
  \right ) \right ] F_{\rm LO}(\tildeptop,\tildeptopbar)
\\
& = \int {\rm d Lips} \left (..., \ptop,\ptopbar,..\right ) \frac{\delta \mt^2}{2\left(\tildemt^2-\ptop \ptopbar\right)}
\left [ 2  + \left(\ptop^\mu-\ptopbar^\mu\right) \left (\frac{\partial }{\partial \ptop^\mu}
  - \frac{\partial }{\partial \ptopbar^\mu}
  \right ) \right ] F_{\rm LO}(\ptop,\ptopbar),
\end{split}
\end{equation}
where in the last  step  we relabelled the momenta  $\tildeptop$ 
and $\tildeptopbar$ back to $p_t$ and $p_{\bar t}$.
While the short-distance masses can be defined in many different ways \cite{tHooft:1973mfk,Czarnecki:1997sz,Beneke:1998rk,Hoang:1999ye,Pineda:2001zq,Hoang:2008yj}, the guiding principle is that they should not contain linear ${\cal O}(\Lambda_{\rm QCD})$ terms.  Therefore, for our purposes, it is sufficient to write 
\begin{equation}
  \mt = {\tilde m}_t  \left ( 1-   \frac{\alpha_s \Cf}{2 \pi} \frac{\pi \lambda }{\mt} \right ),
  \label{eq5.8a}
\end{equation}
so that 
\begin{equation}
\delta \mt = - \mt \; \frac{\alpha_s \Cf}{2 \pi} \frac{\pi \lambda }{\mt},
\;\;\;\; \delta \mt^2 = -2  \mt^2 \; \frac{\alpha_s \Cf}{2 \pi} \frac{\pi \lambda }{\mt}.
\end{equation}

Combining the different terms, we obtain the change  of  the cross section due to the mass shift
\begin{equation}
\begin{split}
  {\rm d} \sigma_{\rm LO}(\mt) - {\rm d} \sigma_{\rm LO}({\tilde m}_t)
  & =  \delta  \sigma^{\rm expl}_{\rm mass} + \delta  \sigma^{\rm impl}_{\rm mass}
  = \frac{\alpha_s \Cf}{2 \pi} \frac{\pi \lambda}{\mt}
  \; \int {\rm d Lips}_{\rm LO}
  \times 
  \\
  & \times \Bigg [
  \frac{\mt^2}{(\ptop \ptopbar-\mt^2)}
\left [ 2  + \left(\ptop^\mu-\ptopbar^\mu\right) \left (D_{t, \mu}
  - D_{\bar{t}, \mu} \right ) \right ] F_{\rm LO}
    \\
    & \;\;\;\;\;\;\; -\left[F_{\rm LO}|_{\hat{\rho}_t = m_t \hatone} -  F_{\rm LO}|_{\hat{\rho}_{\bar t} = m_t \hatone}\right]
  \\ 
   & \;\;\;\;\;\;\; -\mt {\rm Tr} \left [ \hat{\rho}_{t} \left (
     \frac{ \partial\hatN}{\partial \mt} \hat{\rho}_{\bar{t}} \barhatN 
           + \hatN  \hat{\rho}_{\bar{t}} \frac{ \partial \barhatN}{\partial \mt}  \right ) \right ]
      \Bigg  ].
\end{split} \label{eq:massShift}
\end{equation}
Finally, we use  eqs.~(\ref{eqttreal}, \ref{eqttvirt}, \ref{eq:rencontr}, \ref{eq:massShift})
to compute the various ${\cal O}(\lambda)$ 
contributions to the sum of the $tt$ and $\bar t \bar t$ monopoles
and find that the result vanishes 
\be
\delta  \sigma^{\rm expl}_{\rm mass} + \delta  \sigma^{\rm impl}_{\rm mass}
+{\cal T}_\lambda \left [ {\rm d} \sigma_{\rm ren}  \right ]
+
{\cal T}_\lambda \left [ {\rm d} \sigma_V^{t t} + {\rm d} \sigma_V^{ \bar t \bar t} \right ]
+
{\cal T}_\lambda \left [ {\rm d} \sigma_R^{t t} + {\rm d} \sigma_R^{ \bar t \bar t}\right ] = 0.
\ee

As we explained earlier, the ${\cal O}(\lambda)$ contribution to $q \bar q \to t \bar t + X$ cross section can be  calculated as a sum of various dipole and monopole terms. In this section  we have shown that, for each of these terms, the ${\cal O}(\lambda)$ contribution  vanishes. Hence, we conclude that within the renormalon model, there are no ${\cal O}(\Lambda_{\rm QCD})$ corrections to top quark pair production in hadron collisions provided that the leading partonic process is the $q \bar q$ annihilation channel.

\subsection{On the validity of the LBK theorem}
\label{sec:validity}

Recently, in refs.~\cite{Lebiedowicz:2021byo,Lebiedowicz:2023mlz,Lebiedowicz:2023ell} objections were raised about  the validity of the LBK theorem, and one may wonder whether these 
objections have implications for the results 
reported  in this and earlier (e.g. \cite{Makarov:2023ttq}) papers. 
As discussed in~\cite{Lebiedowicz:2021byo,Lebiedowicz:2023mlz,Lebiedowicz:2023ell}, 
potential problems with the derivation of the LBK 
result  stem from the need  to consider the off-shell extensions 
of the Born amplitude, or its extensions to external 
momenta  that do not satisfy momentum
conservation. It is argued in~\cite{Lebiedowicz:2021byo,Lebiedowicz:2023mlz,Lebiedowicz:2023ell} that such extensions  may lead 
to ambiguities  because they  cannot uniquely 
follow from 
amplitudes computed  for external on-shell momenta that satisfy  momentum conservation.
To illustrate this point, we note that if 
we replace  $(p_t+p_{\bar{t}})^2$ either with 
$(p_q + p_{\bar q})^2$ or with $(2m_t^2+2p_t\cdot p_{\bar{t}})$
in the leading order amplitude $F_{\rm LO}^{ij}$,  computations of 
derivatives of $F_{\rm LO}^{ij}$
w.r.t. $t$ or $\bar t$ momentum will yield different 
results. It is therefore important to clarify if and how 
the uniqueness of the result for the radiative 
amplitude is restored. 

To understand this, it is useful  
to realise that the off-shell continuation problem 
and the momentum conservation problem 
have different origins and  resolutions. 
Let us first focus  on the off-shell continuation. 
In this case,
the \emph{final} result~eq.~(\ref{eq2.25}) is \emph{independent} of  any particular off-shell  extension of the amplitude 
squared.
To see this, we note that 
in eq.~(\ref{eq2.25}) terms of the form
$ J_i^\mu J_{j,\mu} F^{ij}_{\rm LO}$ (the $JJ$ terms from now on) as well as terms of the form
$ J_i^\mu L_{i,\mu} F^{ij}_{\rm LO}$ (the $JL$ terms) appear.  
The momenta appearing in the $JJ$ terms are on shell, so they do not depend upon the off-shell continuation.
The $JL$ terms could in principle be affected by the off-shell continuation, but this is not the case,
since the operators $L$ yield zero when applied to the square of the external momenta,
\begin{equation}
  {\bar L}_i^\mu p_i^2 = (J_i^\mu k^\nu {\bar D}_{i,\nu} - {\bar D}_{i}^\mu)p_i^2 \sim J_i^\mu k\cdot p_i- p_i^\mu=0.
\end{equation}
Thus the $L$ derivatives treat the invariants associated with the off-shell extensions 
of external legs as constants, so that,
as far as the derivatives are concerned, working 
with the on-shell $F_{\rm LO}$ functions 
does not affect the result. Therefore, the off-shell continuation of the truncated
Born amplitude is not needed.\footnote{This also follows from  the fact that when expanding 
the amplitude in the off-shellness  
  of  external legs,  one removes  
  denominators of the eikonal currents and  generates   terms that are indistinguishable from the 
  structure-dependent radiation amplitude. Then, the current conservation requirement expresses  
  both the structure-dependent amplitude  \emph{and} 
  the off-shell terms through derivatives of the 
  \emph{on-shell} amplitude.}

On the contrary,  the leading order amplitude  
with external momenta that do not satisfy momentum 
conservation 
is  only introduced for bookkeeping purposes at 
intermediate steps in our construction. 
In fact, we note that in addition to writing 
the expansion of the amplitude squared in the 
small gluon momentum, our computation involves 
a second step where we redefine momenta ($q_t \to p_t $ etc.)  to ensure 
the momentum conservation without the need to account for the  gluon 
momentum and then reexpand the amplitude around these 
conserved momenta values. 
We find that  both the $JJ$ and $JL$ terms are 
affected by the momentum non-conservation 
issue  but, once  $F_{\rm LO}$ is rewritten 
in terms of the conserved momenta, the ambiguities 
must cancel out. 
Hence, we conclude that  our final result for the real-emission contribution to the coefficient of the $\lambda$ term, given by the sum of  eqs.~(\ref{eq:TlRttbar}), (\ref{eq:massShift}) and eq.~(\ref{eq:TlRtq}) with all its variants for $\bar{t}q$, $\bar{t}\bar{q}$ and $t\bar{q}$ dipoles, is not affected by the issues with the LBK theorem pointed out in refs.~\cite{Lebiedowicz:2021byo,Lebiedowicz:2023mlz,Lebiedowicz:2023ell}.

In more detail, any possible contribution of an off-shell 
extension to the on-shell amplitude  disappears 
separately in each   dipole/monopole. 
However, 
the momentum non-conservation extension is more subtle 
since it requires adding
together various 
dipole and monopole contributions. In a particularly simple case of $e^+e^-\to t\bar{t}$,
the cancellation is quite evident, since
only the $t\bar{t}$ dipole and the $tt+\bar{t}\bar{t}$ monopoles contribute, they have the same
colour factor, and the derivative terms are equal and opposite, so that they cancel in the sum.
The case of $q\bar{q}\to t\bar{t}$ is more involved. There we verified that the derivative terms, when acting
on the combination $(p_q+p_{\bar{q}})^2-(p_t+p_{\bar{t}})^2$, sum up to zero, thereby yielding a further check of the correctness of our procedure. 

\section{Kinematic distributions}
\label{sect:kinematics}

We will now study the kinematic distributions in top quark pair production
processes. We consider an observable $X$ that depends exclusively 
on the momentum of the top quark
\begin{equation}
O_X = \int {\rm d} \sigma \; X(\ptop).
\label{eq7.1}
\end{equation}
To compute the ${\cal O}(\lambda)$ contribution to $O_X$, we follow the
approach described  in the previous
sections and write
\be
O_X =
\int {\rm d} \sigma_{\rm LO} \; X(\ptop) +
\int {\rm d} \sigma_{\rm NLO} \; X(\ptop).
\ee
We can write the NLO contribution to the cross section as the sum of
dipoles and monopoles
\be
\int {\rm d} \sigma_{\rm NLO} \; X(\ptop)
  = \sum \limits_{\xa}^{} \int {\rm d}\sigma^{(\xa)}_{\rm NLO} \;
X(\ptop),
\ee
where $\xa$ denotes a particular  dipole or the combination of the $tt$ and $\bar{t}\bar{t}$ monopoles.
In the real-emission contribution of each dipole or monopole, we apply the appropriate momentum mapping defined in Section~\ref{sec:momentummapping} in order to factorise the $k$ integration in the phase space. The difference with respect to the case of the inclusive cross section is the appearance of the observable $X$ in the integrand in eq.~(\ref{eq7.1}). For the real-emission part, we therefore have that
\be
\int {\rm d} \sigma_{R} (q_t,...) \; X(\qtop)
= \sum \limits_{\xa}^{} \int {\rm d} \sigma^{(\xa)}_{R}
(q_t,...) X(p_t)
   + \sum \limits_{\xa}^{} \int {\rm d} \sigma^{(\xa)}_{R}
(q_t,...) \frac{\partial X(p_t)}{\partial p_t^{\mu} }
\delta^{(\xa)}p_t^\mu,
\ee
where $\delta^{(\xa)}p_t^\mu$ is the shift in the top quark momentum given in eq.~(\ref{eq:momentummapping}). Since $\delta^{(\xa)}p_t^\mu \sim \mathcal{O}{(k)}$, one needs, in the second term, ${\rm d} \sigma^{(\xa)}_{R} (q_t,...)$ in the leading soft approximation only.
Hence, we obtain
\be
\int {\rm d} \sigma_{\rm NLO} \; X(\ptop)
= \sum \limits_{\xa}^{} \int {\rm d} \sigma^{(\xa)}_{\rm NLO}
(p_t,...) X(p_t)
   + \sum \limits_{\xa}^{} \int {\rm d} \sigma^{(\xa)}_{R}
(p_t,...) \frac{\partial X(p_t)}{\partial p_t^{\mu} }
\delta^{(\xa)}p_t^\mu,
\ee
where the first term on the right-hand side includes all the terms that
contribute to the calculation of the inclusive cross section
for a particular dipole and monopole except for terms that originate
from the mass redefinition.

The mass redefinition terms affect both the leading order cross section as
well as the observable function $X(\qtop)$ that multiplies it.
Redefining the mass, we obtain
\be
\begin{split}
O_X &=
\int {\rm d} \sigma_{\rm LO} \; X(\ptop)|_{m_t \to {\bar m}_t}
+\sum \limits_{\xa}^{} \int {\rm d} \sigma^{(\xa)}_{\rm NLO}
(p_t,...) X(p_t)
+ \int {\rm d} \sigma_{\rm NLO}^{\rm mass} \; X(p_t)
\\
& + \int {\rm d} \sigma_{\rm LO} \; \frac{\partial X(p_t)}{\partial
p_t^{\mu} } \; \delta^{\rm mass}p_t^\mu
+ \sum \limits_{\xa}^{} \int {\rm d} \sigma^{(\xa)}_{R}
(p_t,...) \frac{\partial X(p_t)}{\partial p_t^{\mu} }
\delta^{(\xa)}p_t^\mu,
\end{split}
\ee
where ${\rm d} \sigma_{\rm NLO}^{\rm mass}$ is the change in the cross
section due to the mass redefinition and $\delta^{\rm mass} p_t$ is the
related shift in the top quark momentum. As was shown in the previous sections,
\be
{\cal T}_{\lambda} \bigg[ \sum \limits_{\xa}^{} \int {\rm d} \sigma^{(\xa)}_{\rm NLO} (p_t,...)
X(p_t)
+ \int {\rm d} \sigma_{\rm NLO}^{\rm mass} \; X(p_t) \bigg] = 0,
\ee
we conclude that
\be
O_X =
{\bar O}_{X}^{\rm  LO}
+ \int {\rm d} \sigma_{\rm LO} \; \frac{\partial X(p_t)}{\partial
p_t^{\mu} } \; \delta^{\rm mass}p_t^\mu
+ \sum \limits_{\xa}^{} \int {\rm d} \sigma^{(\xa)}_{R}
(p_t,...) \frac{\partial X(p_t)}{\partial p_t^{\mu} }
\delta^{(\xa)}p_t^\mu,
\ee
where ${\bar O}_{X}^{\rm  LO}$ is the observable $X$ computed at leading order with the short-distance mass.

In what follows, we will discuss the different contributions to the above
equation. We combine the term that originates from the mass shift with the $tt$ and $\bar t \bar t$ monopole contributions.
The general expression for $\delta^{(\xa)}p_t^\mu$ in eq.~(\ref{eq:momentummapping}) involves the parameters $\alpha$ and $\beta$, and we will specify our choices for them when we discuss the individual dipole and monopole contributions.

\subsection*{Monopoles $tt+\bar t \bar t$}

In this case, we do not need to choose particular  values for  $\alpha$ and $\beta$, and we use the phase-space integrals in Appendix~\ref{app:phaseintegrals} in order to integrate over the gluon momentum $k$. We also combine the mass-redefinition contribution with those of the $tt+\bar t \bar t$ monopoles.\footnote{We remark that for an observable which depends only on $\ptop$, the inclusion of $\bar{t} \bar{t}$ dipole can in principle be avoided. If one uses the alternative treatment of the self-energy contributions to $\bar{t} \bar{t}$ monopole (see Section 7 in ref.~\cite{Makarov:2023ttq}), this monopole does not contribute.} We find
\be
\begin{split}
& {\cal T}_{\lambda} \bigg[ \int {\rm d} \sigma_{\rm LO} \; \frac{\partial X(p_t)}{\partial
p_t^{\mu} } \; \delta^{\rm mass}p_t^\mu
+  \int {\rm d} \sigma^{(tt + \bar t \bar t)}_{R} (p_t,...)
\frac{\partial X(p_t)}{\partial p_t^{\mu} } \delta^{(tt + \bar t \bar
t)}p_t^\mu \bigg]
\\
& =
  - \frac{\alpha_s \Cf}{2 \pi} \frac{\pi \lambda}{\mt} \int {\rm d}
\sigma_{\rm LO} \; \frac{\partial X(\ptop)}{\partial \ptop^\mu }\,
\ptop^\mu.
\end{split}
\ee

\subsection*{Dipole $t\bar{t}$}

For this dipole, we also do not need to specify the $\alpha$ and $\beta$ values. We find
\begin{equation}
\begin{split}
   & {\cal T}_{\lambda} \bigg[ \int {\rm d} \sigma^{(t \bar t) }_{R} (p_t,...) \frac{\partial
X(p_t)}{\partial p_t^{\mu} } \delta^{(t \bar t)}p_t^\mu \bigg]
   \\
   &
   =\frac{\alpha_s}{2 \pi} \frac{\pi \lambda}{\mt} \int {\rm d}
\sigma_{\rm LO}^{t\bar{t}} \; \frac{\partial X(\ptop)}{\partial
\ptop^\mu } \left(\frac{2 (\ptop \ptopbar) \left((\ptop \ptopbar)\,
\ptop^\mu - \mt^2\, \ptopbar^\mu\right)}{(\ptop \ptopbar)^2 - \mt^4}\right).
\end{split}
\end{equation}

\subsection*{Dipole $tq$}

To compute the contribution of this dipole, we take a mapping with
$\alpha = 1$ and $\beta =0$. We then find

\begin{equation}
\begin{split}
& {\cal T}_{\lambda} \bigg[ \int {\rm d} \sigma^{(tq) }_{R} (p_t,...) \frac{\partial
X(p_t)}{\partial p_t^{\mu} } \delta^{(tq)}p_t^\mu \bigg]
=\frac{\alpha_s}{2 \pi} \frac{\pi \lambda}{\mt} \int {\rm d} \sigma_{\rm
LO}^{tq} \; \frac{\partial X(\ptop)}{\partial \ptop^\mu } \left(2
\ptop^\mu - \frac{2\mt^2}{(\ptop p_q)} p_q^\mu \right).
\end{split}
\end{equation}

\subsection*{Remaining dipoles}

In a similar fashion, by choosing corresponding values for
$\alpha$ and $\beta$, one can easily derive similar expressions for the
other dipoles,
\begin{align}
& {\cal T}_{\lambda} \bigg[ \int {\rm d} \sigma^{(t \bar q) }_{ R} (p_t,...) \frac{\partial
X(p_t)}{\partial p_t^{\mu} } \delta^{(t \bar q)}p_t^\mu \bigg]
   =\frac{\alpha_s}{2 \pi} \frac{\pi \lambda}{\mt} \int {\rm d}
\sigma_{\rm LO}^{t \bar q} \; \frac{\partial X(\ptop)}{\partial
\ptop^\mu } \left(-2 \ptop^\mu + \frac{2\mt^2}{(\ptop p_{\bar q})} p_{\bar q}^\mu \right),
\\
& {\cal T}_{\lambda} \bigg[ \int {\rm d} \sigma^{(\bar t  q) }_{ R} (p_t,...) \frac{\partial
X(p_t)}{\partial p_t^{\mu} } \delta^{(\bar t q)}p_t^\mu \bigg] = 0,
\\
& {\cal T}_{\lambda} \bigg[ \int {\rm d} \sigma^{(\bar t  \bar q) }_{ R} (p_t,...)
\frac{\partial X(p_t)}{\partial p_t^{\mu} } \delta^{(\bar{t} \bar
{q})}p_t^\mu \bigg] = 0.
\end{align}

\subsection*{Linear shift in the observable distributions}
We now combine the results derived for the individual dipoles and monopoles. It is easy to see that for processes that have the same colour structure as $q{\bar q}\to {t\bar{t}}$, we can express the colour-correlated cross section through combinations of Casimir invariants and the leading order cross section.
We then find 
\begin{equation}
  {\rm d} \sigma_{\rm LO}^{\xa} 
= C^{\xa} \; {\rm d} \sigma_{\rm LO},
\end{equation}
where coefficients  $C^{\xa}$ are dipole-specific colour factors.  They read 
   \begin{equation}
   \begin{split}
       &C^{tt}= C^{\bar t \bar t} = C_F,  \hspace{3.15cm} C^{t \bar t} = C_F - C_A/2, 
       \\
       &C^{tq}= C^{ \bar{t} \bar{q} } = 2 \Cf -  \Ca/2,  \hspace{1.5cm}     C^{\bar{t} q}= C^{ t \bar{q} } = 2 \Cf - \Ca,
   \end{split}
   \label{eq:colourqqbar}
   \end{equation}

Using these colour factors, we write the expression for the observable in the following way
\begin{equation}
   \begin{split}
     O_X &  = \int {\rm d} \sigma_{\rm LO} \left [ X(\ptop) |_{m_t \to
{\bar m}_t}
       + \frac{\alpha_s}{2 \pi }
       \frac{\pi \lambda}{\mt} \left(\sum_{\xa} C^{\xa}\,
l^\mu_{\xa}\right) \frac{\partial X( \ptop)}{\partial \ptop^\mu}
       \right ],
   \end{split}
   \end{equation}
where the momenta  $l_{\xa}$ can be extracted from the results derived in the previous subsection,
\be
     l^\mu_{\xa} =
\begin{cases}
  - \ptop^\mu, & \text{for } (\xa) = (tt + \bar{t}\bar{t}),
  \\    
2 (\ptop \ptopbar) \left((\ptop \ptopbar)\,
\ptop^\mu - \mt^2\, \ptopbar^\mu\right)/\left((\ptop \ptopbar)^2 - \mt^4 \right),    & \text{for } (\xa) = (t \bar{t}),
\\ 
2 \ptop^\mu - 2\mt^2\, p_q^\mu / (\ptop p_q),    & \text{for } (\xa) = (t q),
\\ 
-2 \ptop^\mu + 2\mt^2\, p_{\bar q}^\mu /(\ptop p_{\bar q}),    & \text{for } (\xa) = (t \bar{q}),
\\
0, & \text{for } (\xa) = (\bar{t} q) \text{ and }  (\bar{t} \bar{q}).
\end{cases}
\ee

     It is clear that the above result can be written as the shift in
the argument of the function $X$. We find
     \be
{\cal O}_X
      = \int {\rm d} \sigma_{\rm LO} \;  X\left(\ptop+ \frac{\alpha_s}{2
\pi }
      \sum_{\xa} C^{\xa}\, \delta p_{t, \xa} \right),
\ee
where
\be
\delta  p_{t, \xa} =\frac{\pi \lambda}{\mt}\;l_{\xa}.
\ee

In a similar fashion, one can derive the corresponding expressions for observables that depend on the momentum of the anti-top. We provide the complete expressions in Appendix~\ref{appObservablesX}.

\section{Applications to simple  kinematic distributions}
\label{sec:applications}
In this section we compute the linear power corrections to three simple observables -- the
top quark transverse momentum,  the top quark rapidity and the $t \bar t$ invariant mass -- focusing on the process $q \bar{q} \to t \bar{t}$ with no additional colour-neutral particles in the final state.  Complete formulas for other  processes e.g. 
$q \bar{q} \to t \bar{t} +X$
 and  $e^+ e^- \to t \bar{t} + X$ are given in Appendix~\ref{appObservablesX}.

Before we begin, we recall that, in the large $\nf$ framework, the term linear in $\lambda$ in the cross sections is related to the leading factorial growth of the coefficients of the perturbative expansion in $\as$, according to the formula reported, for example, in Appendix A of ref.~\cite{FerrarioRavasio:2020guj}. From that reference (following the same notation) it is also clear that the renormalon ambiguity is obtained by replacing  $\as\lambda$ with a non-perturbative scale of order $\Lambda_{\text{QCD}}$. For the following estimates, we will only need to know that
we should replace $\as\lambda$ with a scale parameter, that we will fix according to current estimates of the renormalon ambiguity in the on-shell top mass.

The well-known expressions for the top quark transverse momentum, its rapidity in the partonic center-of-mass frame and the $t \bar t$ invariant mass read
\begin{equation}
  {\ptop}_\perp  =\sqrt{\ptop^\mu g_{\perp,\mu \nu} \ptop^\nu},
  \;\;\;\;\;\;\;
   y_t = \frac{1}{2} \ln \frac{p_{\bar{q}} \ptop}{p_q \ptop}, 
   \;\;\;\;\;\;\;
   s_{t \bar t} = (p_t + p_{\bar t})^2,
   \label{eq:definitionkinobs}
\end{equation}
where
\begin{equation}
  g_{\perp}^{\mu \nu} = \frac{p_q^\mu p_{\bar{q}}^\nu + p_{\bar{q}}^\mu p_q^\nu}{p_q p_{\bar{q}}}-g^{\mu \nu}.
\end{equation}

Applying the formalism of Section~\ref{sect:kinematics} and defining $ \tau= 4 m_t^2 / s_{t \bar{t}}$, we find
\begin{align}
   \frac{ \delta_{\rm NP} \left [{\ptop}_{\perp}  \right ]
}{{\ptop}_{\perp}}
  =&\frac{\alpha_s}{2 \pi } \frac{\pi \lambda}{\mt}\; \frac{(2 C_F - C_A \tau)}{2(1-\tau)},
  \\[10pt]
   \delta_{\rm NP} \left [ y_t \right ]
   =& \frac{\alpha_s}{2 \pi } \frac{\pi \lambda}{\mt}\; \left[\left(3
\Ca-8 \Cf\right) \tau \cosh^2{y_t}
     - \left(\Ca-2\Cf\right) \frac{\tau (2 - \tau)}{4(1 - \tau)} \sinh{(2 y_t)}\right],
 \label{eq8.4}
    \\[10pt]
   \frac{\delta_{\rm NP} \left [ s_{t \bar t} \right ]}{s_{t \bar{t}}}
   =& \frac{\alpha_s}{2 \pi } \frac{\pi \lambda}{\mt}\; \bigg[ 2 \Cf\, (1-\tau) - \Ca\, \tau \cosh{(2 y_t)} + \left( 3 \Ca - 8 \Cf \right) \tau \sinh{(2 y_t)} \bigg].
   \label{eq8.5}
\end{align}

\begin{figure}
    \centering
    
        \includegraphics[width=0.5\textwidth]{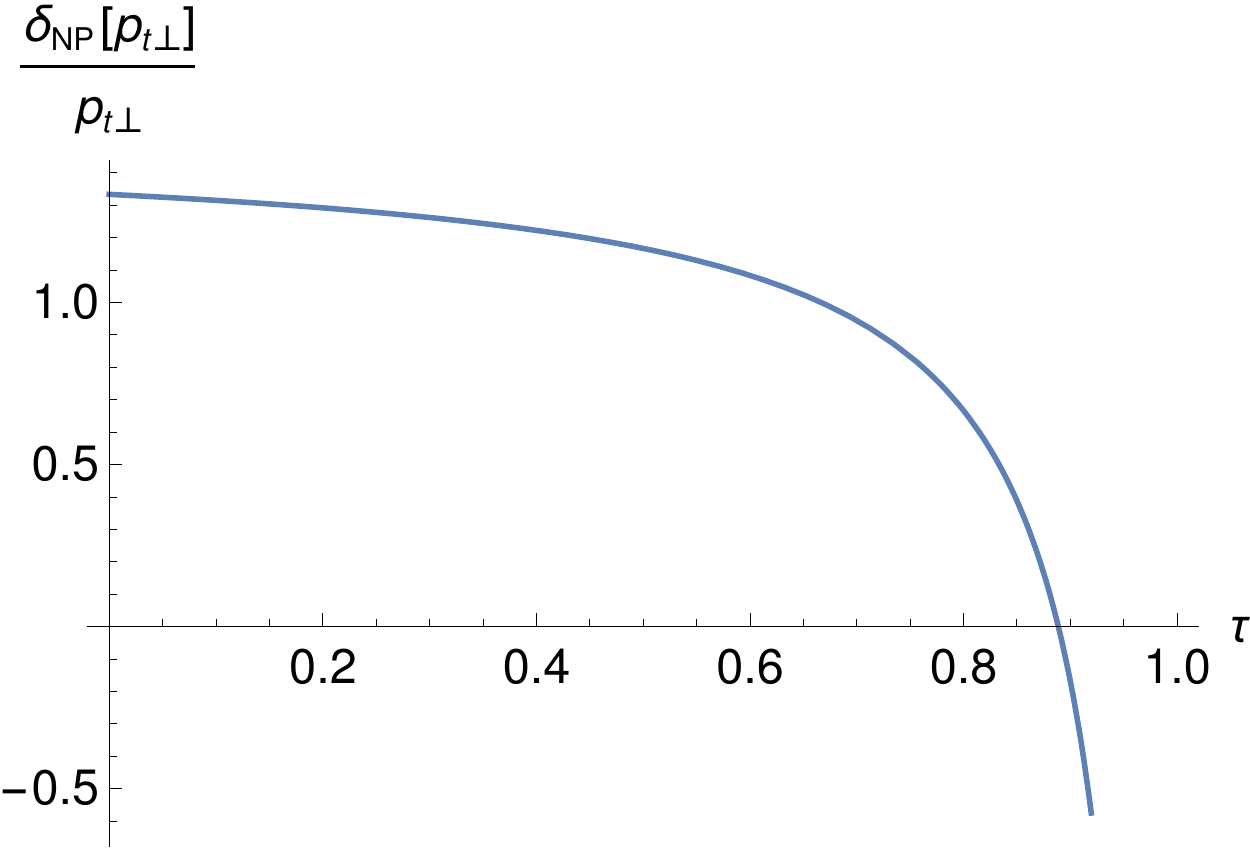}
    
    \caption{Plot of $\delta_{\rm NP} \left [{\ptop}_{\perp}  \right ]
/{\ptop}_{\perp}$ as function of $\tau$. The global factor of $\alpha_s/(2 \pi)\, \pi \lambda/\mt$ has been set to one.}
    \label{figshifts_pt}
\end{figure}

\begin{figure}
    \centering
    
    \includegraphics[width=0.6\textwidth]{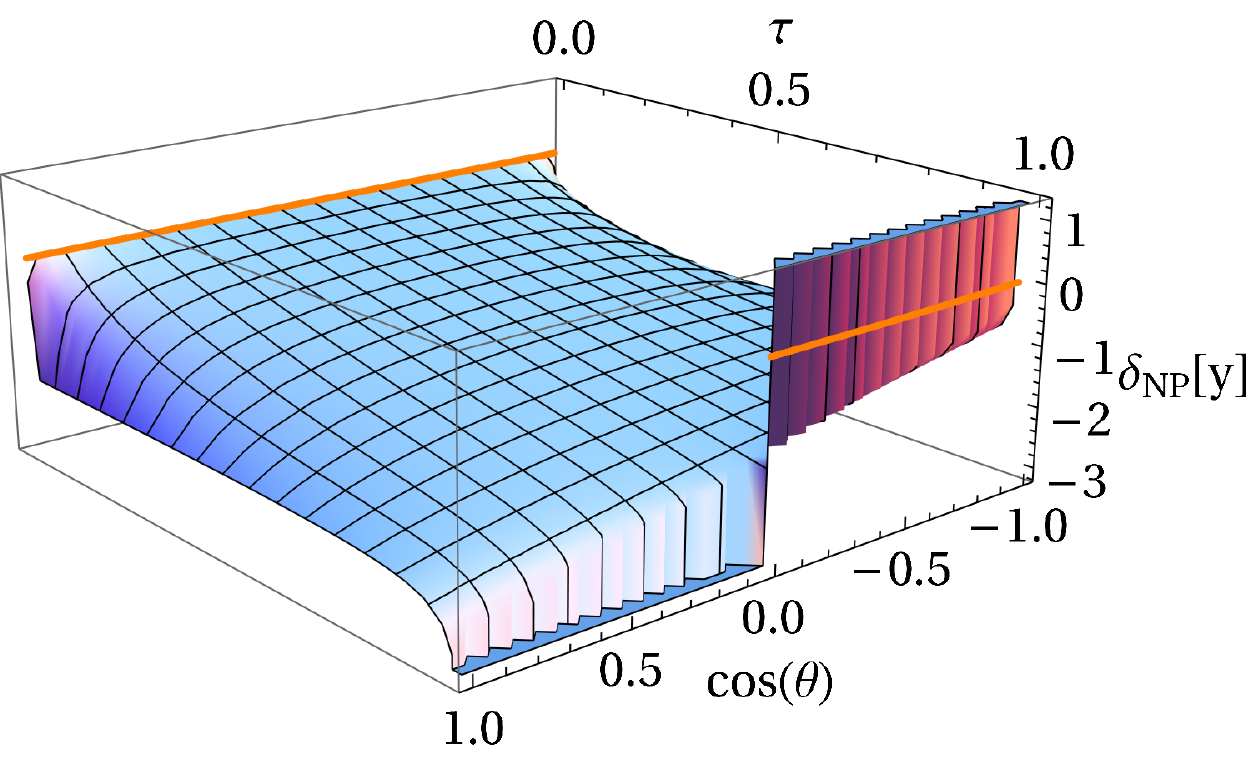}

    \caption{Plot of $\delta_{\rm NP} \left [ y_t \right ]$ as function of $\tau$ and $\cos{\theta}$ (see text for details). The global factor of $\alpha_s/(2 \pi)\, \pi \lambda/\mt$ has been set to one. The orange lines indicate the intersection with the plane of vanishing shift.}
\label{figshifts_y}
\end{figure}

\begin{figure}
    \centering
    
    \includegraphics[width=0.6\textwidth]{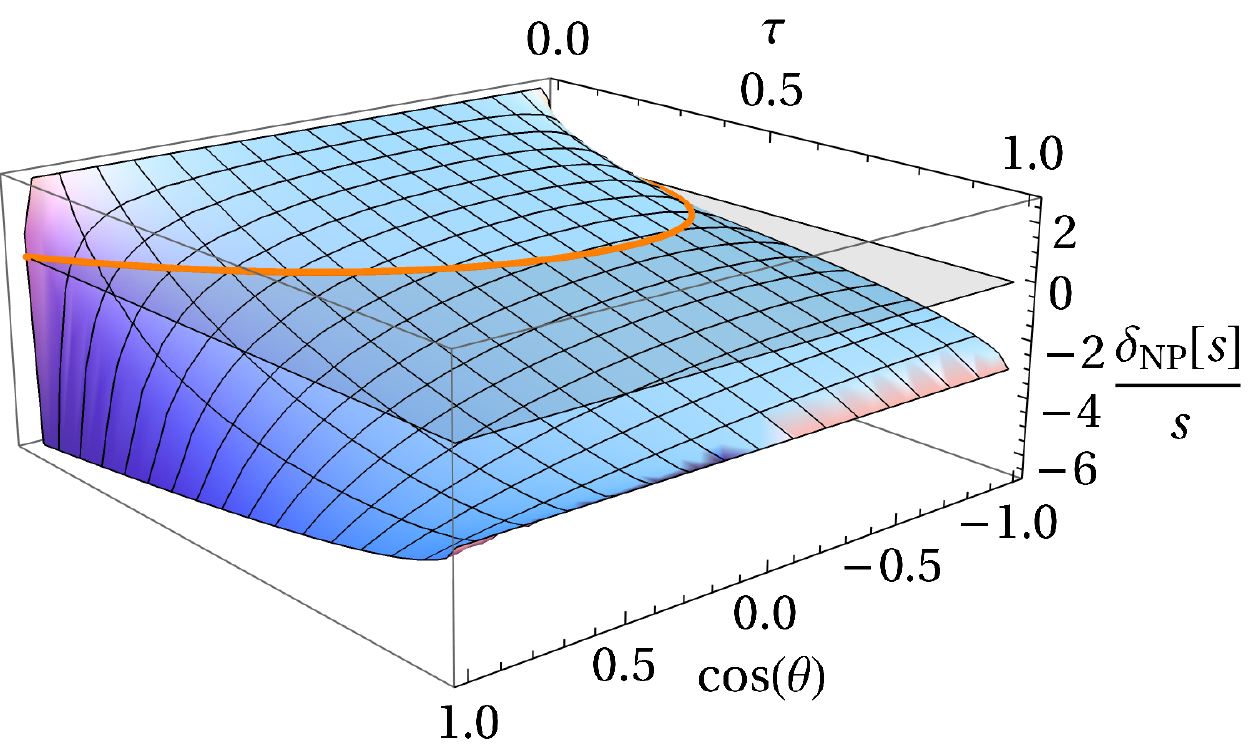}
    
    \caption{Plot of $\delta_{\rm NP} \left [ s_{t \bar t} \right ]/s_{t \bar{t}}$ shift as function of $\tau$ and $\cos{\theta}$ (see text for details). The global factor of $\alpha_s/(2 \pi)\, \pi \lambda/\mt$ has been set to one. In the plot, we have added a transparent plane of vanishing shift. The orange lines indicate the intersection with the plane of vanishing shift.}
    \label{figshifts_s}
\end{figure}

Interestingly, these shifts exhibit non-trivial dependencies on the QCD colour factors and on the kinematics of the underlying 
$q \bar q \to t \bar t $ process.\footnote{We note that 
very close to the threshold, 
fixed-order perturbative computations break down. This means that our results in this region should 
be interpreted
with care.} To visualise them, we display the shifts in Figs.~\ref{figshifts_pt} - \ref{figshifts_s}.
We observe that the transverse momentum shift is large and negative around the partonic threshold and that the sign is driven by the non-Abelian Casimir $C_A$.  The transverse momentum shift changes the  sign at 
\be
\sqrt{s} = 2 m_t \sqrt{\frac{C_A}{2 C_F}},
\ee
which, numerically, is $\mathcal{O}{(20)}~{\rm GeV}$ above the $t \bar t$ threshold. At larger invariant masses, the non-perturbative shift is dominated by the ``Abelian'' contribution proportional to $C_F$.

The shifts in $y_t$ and $s_{t \bar t}$ depend on both the invariant mass of the $t \bar t$ pair and the rapidity of the top quark. Since we work in the partonic center-of-mass frame, it is convenient to express the rapidity of the top quark through the scattering angle $\theta$ of $t$ relative to $q$ using 
\begin{equation}
    y_t=\frac{1}{2}\log{\left(\frac{1+\sqrt{1-\tau} \cos{\theta}}{1-\sqrt{1-\tau} \cos{\theta}}\right)}.
\end{equation}
Hence, to visualise the shifts in $y_t$ and $s_{t \bar t}$, we use two-dimensional plots in $\tau$ and $\cos{\theta}$, see Figs.~\ref{figshifts_y} - \ref{figshifts_s}.

A peculiar  feature of these shifts is that they induce forward-backward asymmetry in 
$t \bar t$ production. This is obvious from the presence of $\sinh (2 y_t)$ terms in eqs.~(\ref{eq8.4}, \ref{eq8.5}). Moreover, these $y_t$-odd shifts are again enhanced in the threshold region. To see this, we expand eq.~(\ref{eq8.4}) around threshold, $\tau=1$, and find 
\be
\lim_{\tau \to 1} \delta_{\rm NP}[y_t]
 = - \frac{\alpha_s }{2\pi} \frac{\pi \lambda}{m_t}\;
 \frac{ (C_A - 2 C_F) }{2(1-\tau)}\; y_t.
\ee
Comparing this shift with the shift of ${\ptop}_{\perp}$ in the threshold region, we observe that the relative shifts are, in fact, identical and determined by the same colour factors involving both $C_F$ and $C_A$,
\begin{equation}
    \lim_{\tau \to 1} \frac{\delta_{\rm NP}[y_t]}{y_t} = \lim_{\tau \to 1} \frac{ \delta_{\rm NP} \left [{\ptop}_{\perp}  \right ]}{{\ptop}_{\perp}}.
\end{equation}
In contrast to this, the relative shift for the $t\bar t$ invariant mass in the threshold region 
is constant and involves only the non-Abelian colour factor,
\begin{equation}
    \lim_{\tau \to 1} \frac{\delta_{\rm NP} \left [ s_{t \bar t} \right ]}{s_{t \bar{t}}} = - \frac{\alpha_s C_A}{2 \pi } \frac{\pi \lambda}{\mt}.
\end{equation}

In the opposite $ \tau =0 $ limit which correspond 
to the high-energy regime,  we note that, while the shift in $y_t$ vanishes, the relative shifts of ${\ptop}_{\perp}$ and $s_{t \bar t}$ are purely ``Abelian'' and can be related to the shift in the mass redefinition as follows
\begin{equation}
    \frac{ \delta_{\rm NP} \left [\mt  \right ]}{\mt} = \lim_{\tau \to 0} \frac{ \delta_{\rm NP} \left [{\ptop}_{\perp}  \right ]}{{\ptop}_{\perp}} = \frac{1}{2} \lim_{\tau \to 0} \frac{\delta_{\rm NP} \left [ s_{t \bar t} \right ]}{s_{t \bar{t}}}=\frac{\alpha_s C_F}{2 \pi } \frac{\pi \lambda}{\mt}.
\end{equation}

\begin{figure}[t]
    \centering
    \includegraphics[width=0.49\textwidth]{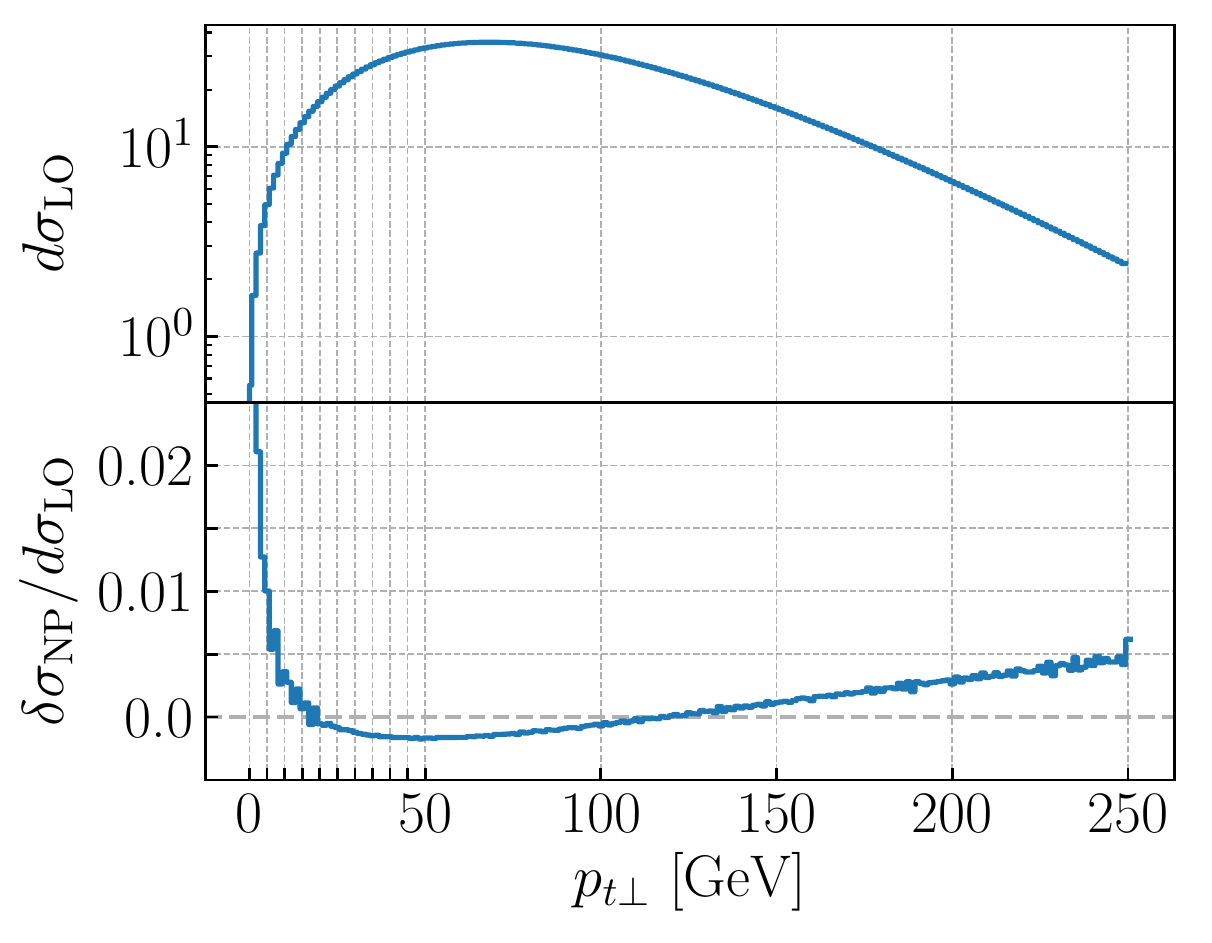} \includegraphics[width=0.5\textwidth]{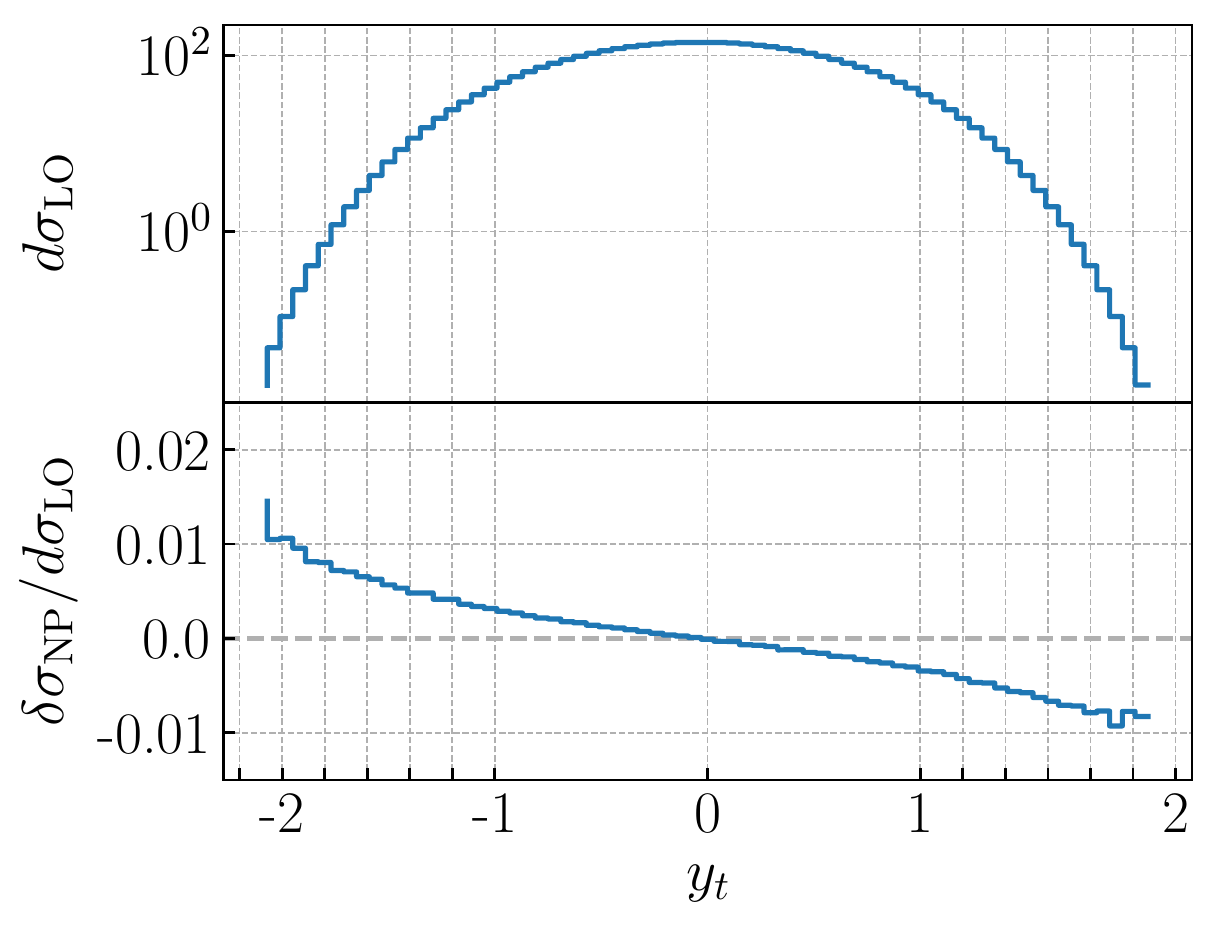}
\includegraphics[width=0.49\textwidth]{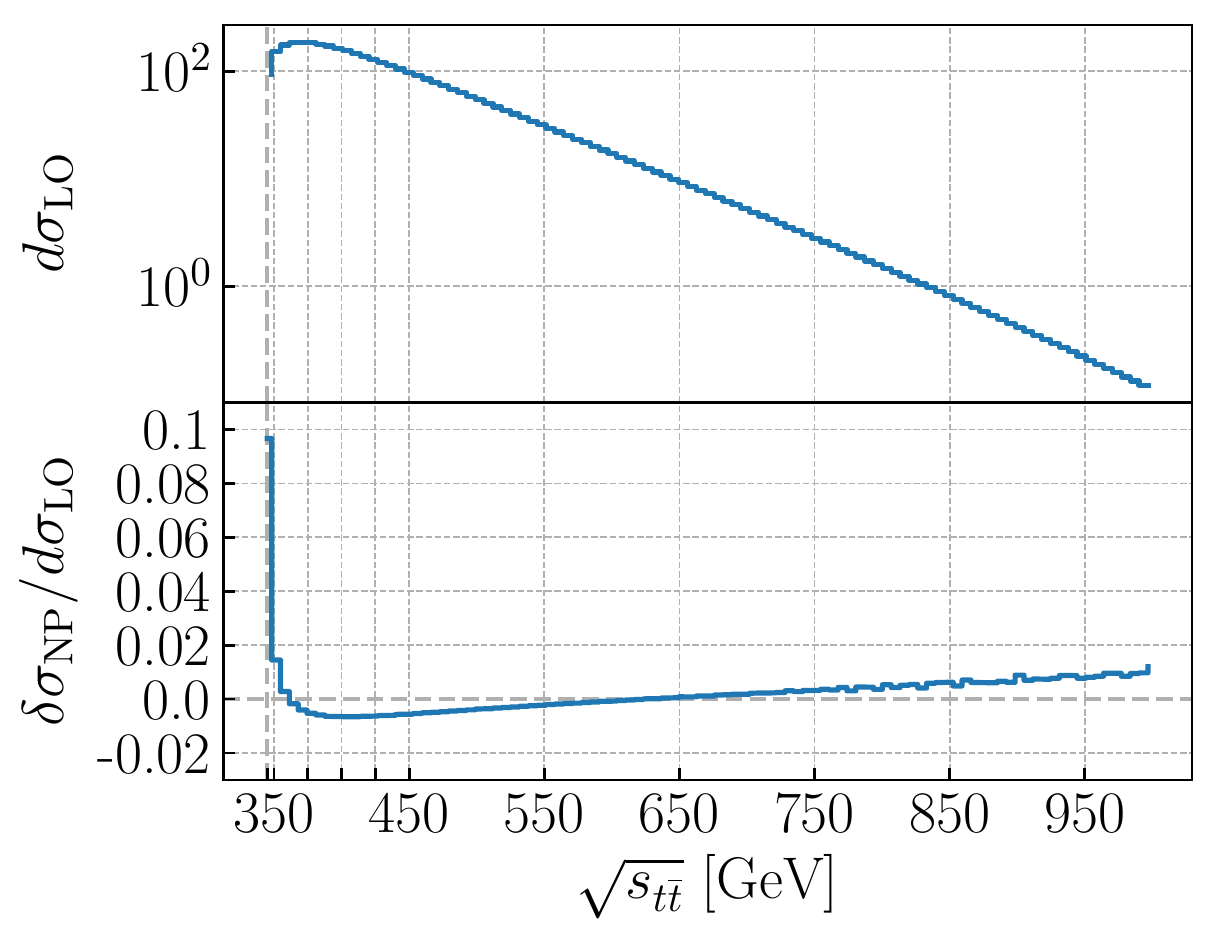}
    
    \caption{Non-perturbative shifts in top quark transverse momentum, lab-frame rapidity and $t \bar t$ invariant mass distributions at the Tevatron for the $q \bar q \to t \bar t$ process. The center-of-mass energy is set to $\sqrt{s}=1.8~{\rm TeV}$. The upper pane shows the leading order distribution. The lower pane shows the ratio 
    $\delta\sigma_\mathrm{NP}/{\rm d} \sigma_{\rm LO} = [ {\rm d} \sigma_{\rm LO}(v+\delta v_{\rm NP}) - {\rm d} \sigma_\mathrm{LO}(v)]/{\rm d} \sigma_{\rm LO}(v)$ for an observable $v$ affected by a non-perturbative 
    shift $\delta v_{\rm NP}$. 
    See text for details.}
    \label{diffsigma_shift}
\end{figure}

\vspace*{0.3cm}
We have also computed the non-perturbative shifts for  
basic top-quark kinematic distributions in the $p \bar p \to t \bar t$ process at the Tevatron; the results  are shown in Fig.~\ref{diffsigma_shift}. To  assign a numerical value to the product of $\alpha_s$ and the gluon mass $\lambda$, we assume that the non-perturbative shift in the value of the top quark pole mass is $200~{\rm MeV}$ \cite{Beneke:2016cbu, Hoang:2017btd, Schwienhorst:2022yqu}. Then, using eq.~(\ref{eq5.8a}) we obtain 
\be
\alpha_s \lambda 
 =   \frac{ 0.4~{\rm GeV}}{C_F} = 0.3~{\rm GeV}.
\ee
Furthermore, we employ the central value of the NNPDF31\_lo\_as\_0118 parton distribution function \cite{NNPDF:2017mvq}, take $m_t = 172.5~{\rm GeV}$ and set the factorisation and the renormalisation scales to $\mu_F = \mu_R = \mt$.\footnote{The numerical value of the top quark mass is chosen for the illustration purposes only. In principle, as we mentioned several times in the text, we must use a short-distance top quark mass to ensure that ${\cal O}(\Lambda_{\rm QCD})$ corrections to the total cross section vanish.} 

We observe (c.f. Fig.~\ref{diffsigma_shift}) that non-perturbative corrections in ${\ptop}_{\perp}$ and $s_{t \bar t}$ distributions can be significant in the corresponding threshold regions.
Although in  ${\ptop}_{\perp}$ distribution large effects are confined to a region which ends about  $5~{\rm GeV}$ above the ${\ptop}_{\perp}$-threshold, for the $t \bar t$ invariant mass distribution ${\cal O}(1 \%)$ effects appear in a broader interval of the invariant masses that extends to about  $450~{\rm GeV}$.  Non-perturbative corrections to the 
rapidity distribution  are small at central rapidities   but  become larger at $|y_t| > 1.5$ where the leading order rapidity distribution starts to decrease rapidly.

\section{Conclusions}
\label{sect:concl}

In this paper we computed linear non-perturbative 
${\cal O}(\Lambda_{\rm QCD})$
corrections to top quark pair  production 
in hadron collisions under 
the assumption that $q \bar q \to t \bar t $ is the dominant partonic channel. 
Our starting point is the renormalon model.
Traditionally, the renormalon calculus is used to compute linear power corrections to processes
without gluons at the tree level, which is clearly not the case for the $t \bar t$ production in hadron collisions.
However, we have argued that, for \emph{quark initiated} partonic processes, i.e. for $q \bar q \to t \bar t$,
the renormalon calculus is still applicable, because of the large virtuality of the gluon in the Born diagram.

We have shown how to compute the linear power corrections efficiently using a generalisation of the Low-Burnett-Kroll theorem to processes with colour charges.
In this case, the first subleading soft corrections can be written in terms of colour-correlated matrix elements, in a form that exhibits the dipole structure typical of soft radiation.  We have further shown that, for inclusive total cross sections expressed through a short-distance top quark mass, the ${\cal O}(\Lambda_{\rm QCD})$ contributions vanish and, if a proper mapping of momenta is chosen, this occurs separately for each of the colour dipoles.  

Finally, we studied the non-perturbative corrections to kinematic distributions that depend on the momenta of the top and anti-top quarks. Our formalism allows us to compute them in a straightforward manner.  Although these are not particularly large numerically, they exhibit interesting dependencies on the  kinematics of the Born process and on the QCD colour factors.  For example, the relative correction to the transverse momentum distribution of the top, ${\ptop}_{\perp}$, is large and negative close to the $t \bar t$ threshold, where the sign is driven by the non-Abelian colour factor $C_A$.  However, the sign changes at $\sqrt{s_{t\bar{t}}} =2 m_t \, \sqrt{C_A/(2 C_F)}$ which is about $20~{\rm GeV}$ above the $t \bar t$ threshold.  Furthermore, the ${\cal O}(\Lambda_{\rm QCD})$ corrections to the top quark rapidity distribution induce forward-backward asymmetry, which is particularly enhanced in the threshold region.  Hence, it appears from our analysis that even for a relatively simple $2 \to 2$ process that we consider here, the renormalon model predicts interesting kinematic dependencies of non-perturbative power suppressed effects that relate to such fundamental properties of QCD as gluon self-interactions.

As the last observation, we notice that both in the single top production case discussed in ref.~\cite{Makarov:2023ttq} and in the present case, no linear power corrections are present in the inclusive total cross section if one uses a short-distance mass scheme. Although the full analysis of hadronic $t \bar t$ production that incorporates the $gg$ partonic channel remains an outstanding task,  these persistent cancellations  hint at the possibility that this property holds in general. 
Assuming that this is the case, this would imply that  short-distance mass schemes (e.g.  the ${\overline {\rm MS}}$ scheme \cite{tHooft:1973mfk}, or schemes of refs.~\cite{Czarnecki:1997sz,Beneke:1998rk,Hoang:1999ye,Pineda:2001zq,Hoang:2008yj}) are preferable for computing the total cross section and, if a heavy quark mass parameter is extracted from the cross-section measurement, the quoted result should be in a  chosen short-distance mass scheme.

\section*{Acknowledgments}
The research of K.M. and S.M. was supported by the German Research Foundation (DFG, Deutsche Forschungsgemeinschaft) under grant 396021762-TRR 257.
P.~N. acknowledges the support of the Humboldt foundation.

\appendix

\section{Loop and real-emission integrals required for computing linear power corrections}\label{app:integrals}
   
In this Appendix we give the results for the phase-space and loop integrals that occur in the real emission and virtual contributions respectively. In order to present the results in a compact form, we make use of the variable
\begin{equation}
\delta = \frac{1}{(2\pi)^2} \frac{\lambda \pi}{\mt}.
\end{equation}

\subsection{Real emission integrals}
\label{app:phaseintegrals}

The phase-space integrals required for computing the real-emission contribution to top quark pair production read\footnote{We only
  display ${\cal O}(\lambda)$ contributions to these integrals.}
\begin{align}
  & I_1 = {\cal T}_\lambda \left [  \int
    \frac{{\rm d}^4 k}{(2\pi)^3} \delta_+(k^2 - \lambda^2) \;  \frac{1}{(2 \ptop k)}
    \right ]
    = -\frac{\delta}{4},
  \\
  & I_2 = {\cal T}_\lambda \left [  \int
    \frac{{\rm d}^4 k}{(2\pi)^3} \delta_+(k^2 - \lambda^2) \;  \frac{k^{\mu}}{(2 \ptop k)^2}
    \right ]
    = -\frac{\delta}{8} \frac{1}{\mt^2} \; \ptop^\mu,
  \\
  & I_3 = {\cal T}_\lambda \left [  \int
    \frac{{\rm d}^4 k}{(2\pi)^3} \delta_+(k^2 - \lambda^2) \;  \frac{k^{\mu}k^{\nu}}{(2 \ptop k)^3}
    \right ]
    = \frac{\delta}{32} \frac{1}{\mt^2} \left(g^{\mu \nu} - \frac{3}{\mt^2} \; \ptop^\mu \ptop^\nu \right),
    \\
    & I_4 = {\cal T}_\lambda \left [  \int
    \frac{{\rm d}^4 k}{(2\pi)^3} \delta_+(k^2 - \lambda^2) \;  \frac{k^{\mu}}{(2 \ptop k) (2 \ptopbar k)}
    \right ]    = -\frac{\delta}{8}\; \frac{1}{(\ptop \ptopbar)+\mt^2} \left(\ptop^\mu + \ptopbar^\mu\right),
    \\
    & I_5 = {\cal T}_\lambda \left [  \int
    \frac{{\rm d}^4 k}{(2\pi)^3} \delta_+(k^2 - \lambda^2) \;  \frac{\lambda^2}{(2 \ptop k)^2 (2 \ptopbar k)}
    \right ]
    = \frac{\delta}{16}\; \frac{1}{(\ptop \ptopbar)+\mt^2}.
  \\
  & I_6 = {\cal T}_\lambda \left [  \int \frac{{\rm d}^4 k}{(2\pi)^3} \delta_+(k^2 - \lambda^2) \;
    \frac{ k^\mu }{(2 \ptop k) (-2 p_q k)} 
    \right ]
    = \frac{\delta}{8} \frac{1}{( \ptop p_q)} \left ( \ptop^\mu - \frac{\mt^2 }{ (\ptop p_q)} \; p_q^\mu \right ).
  \\
  & I_7 = {\cal T}_\lambda \left [  \int \frac{{\rm d}^4 k}{(2\pi)^3} \delta_+(k^2 - \lambda^2) \;
    \frac{\lambda^2 }{(2 \ptop k)^2 (-2 p_q k)} 
    \right ]
    = - \frac{\delta}{16} \frac{1}{(  \ptop p_q) },
  \\
  & I_8 = {\cal T}_\lambda \left [  \int \frac{{\rm d}^4 k}{(2\pi)^3} \delta_+(k^2 - \lambda^2) \;
    \frac{\lambda^2 }{(2 \ptop k) (-2 p_q k)^2} 
    \right ]
    = - \frac{\delta}{16} \frac{\mt^2 }{( \ptop p_q)^2}.
\end{align}

\subsection{Loop integrals}
\label{apploopi}
The required loop integrals read
\begin{align}
& V_{1}={\cal T}_\lambda \left [ 
    -i \int \frac{{\rm d}^4 k}{(2\pi)^4} \frac{1}{(k^2 - \lambda^2)}
    \frac{1}{\left(2p_t k\right)} \right ] = -\frac{\delta}{4},
\\
\begin{split}
  & V_{2}={\cal T}_\lambda \left [ 
    -i \int \frac{{\rm d}^4 k}{(2\pi)^4} \frac{1}{(k^2 - \lambda^2)}
    \frac{k^\mu}{\left(2p_t k\right) \left(-2p_{\bar t} k\right)} \right ] = \frac{\delta}{8} \frac{1}{(p_t p_{\bar t})-m_t^2} \left(p_t^\mu - p_{\bar t}^\mu\right),
  \end{split}
\\
  & V_{3}={\cal T}_\lambda \left [ 
    -i \int \frac{{\rm d}^4 k}{(2\pi)^4} \frac{1}{(k^2 - \lambda^2)}
    \frac{\lambda^2}{\left(2p_t k\right) \left(-2p_{\bar t} k\right)^2} \right ] = - \frac{\delta}{16} \frac{1}{ (p_t p_{\bar t} )-m_t^2}.
    \\
  & V_4 = {\cal T}_\lambda \left [ 
    -i \int \frac{{\rm d}^4 k}{(2\pi)^4} \frac{1}{(k^2 - \lambda^2)} \frac{ k^\mu }{(2 \ptop k) (2 p_q k)} 
    \right ]
    = -\frac{\delta}{8} \frac{1}{( \ptop p_q)} \left ( \ptop^\mu - \frac{\mt^2 }{ (\ptop p_q)} \; p_q^\mu \right ).
  \\
  & V_5 = {\cal T}_\lambda \left [ 
    -i \int \frac{{\rm d}^4 k}{(2\pi)^4} \frac{1}{(k^2 - \lambda^2)} \frac{\lambda^2 }{(2 \ptop k)^2 (2 p_q k)} 
    \right ]
    = \frac{\delta}{16} \frac{1}{(  \ptop p_q) },
  \\
  & V_6 = {\cal T}_\lambda \left [ 
    -i \int \frac{{\rm d}^4 k}{(2\pi)^4} \frac{1}{(k^2 - \lambda^2)} \frac{\lambda^2 }{(2 \ptop k) (2 p_q k)^2} 
    \right ]
    = - \frac{\delta}{16} \frac{\mt^2 }{( \ptop p_q)^2}.
\end{align}

\section{Observables for a more general process of $q \bar{q} \to t \bar{t} + X$}
\label{appObservablesX}

In this Appendix we give the relevant expressions for observables in a more general setting $q \bar{q} \to t \bar{t} +X$ and also briefly discuss the case for $e^+ e^- \to t \bar{t} +X$. In the presence of $X$, the symmetry between $t$ and $\bar{t}$ breaks down and, hence, we need to consider this case explicitly.

For observables depending on the $\bar{t}$ momentum, we acquire the shifts
\begin{equation}
\bar{l}^\mu_{\xa} =
\begin{cases}
  - \ptopbar^\mu, & \text{for } (\xa) = (tt + \bar{t}\bar{t}),
  \\    
2 (\ptop \ptopbar) \left((\ptop \ptopbar)\,
\ptopbar^\mu - \mt^2\, \ptop^\mu\right)/\left((\ptop \ptopbar)^2 - \mt^4 \right),    & \text{for } (\xa) = (t \bar{t}),
\\ 
-2 \ptopbar^\mu + 2\mt^2\, p_q^\mu / (\ptopbar p_q),    & \text{for } (\xa) = (\bar{t} q),
\\ 
2 \ptopbar^\mu - 2\mt^2\, p_{\bar q}^\mu /(\ptopbar p_{\bar q}),    & \text{for } (\xa) = (\bar{t} \bar{q}),
\\
0, & \text{for } (\xa) = (t q) \text{ and }  (t \bar{q}).
\end{cases}
\end{equation}
Using this, for general observables that depend on both momenta $p_t$, $p_{\bar t}$ and the top mass $m_t$, the linear shift reads
\begin{equation}
\begin{split}
  &{\cal T}_\lambda [O_X] = \frac{\alpha_s }{2 \pi } \frac{\pi \lambda}{\mt} \; \int {\rm d} \sigma_{\rm LO} \; \Bigg[ \left(\sum_{\xa} C^{\xa}\, l^\mu_{\xa}\right) \frac{\partial X(\ptop,\ptopbar,\mt^2)}{\partial \ptop^\mu }
  \\
  & \hspace{3.7cm} + \left(\sum_{\xa} C^{\xa}\, \bar{l}^\mu_{\xa}\right) \frac{\partial X(\ptop,\ptopbar,\mt^2)}{\partial \ptopbar^\mu }-\mt \frac{\partial X(\ptop,\ptopbar,\mt^2)}{\partial \mt} \Bigg].
  \end{split}
\end{equation}

In the following, we consider the same observables as in Section~\ref{sec:applications} for $q\bar{q} \to t \bar{t} + X$, where $s = (p_q + p_{\bar q})^2 \neq  s_{t \bar{t}}= (\ptop + \ptopbar)^2$. We will give the split-down for the different monopole and dipole contributions and only explicitly insert the colour coefficients for the monopoles, the remaining coefficients can be extracted from eq.~(\ref{eq:colourqqbar}). The definitions for ${\ptopbar}_\perp$ and $y_{\bar t}$ follow the equivalent definitions for ${\ptop}_\perp$ and $y_{t}$ in eq.~(\ref{eq:definitionkinobs}) respectively, but with $\ptop$ replaced by $\ptopbar$ instead.

The shift in the transverse momenta for the top and anti-top read,
\begin{align}
\begin{split}
  \frac{ \delta_{\rm NP} \left [{\ptop}_{\perp}  \right ] }{{\ptop}_{\perp}} =& \frac{\alpha_s}{2 \pi }
  \frac{\pi \lambda}{\mt}\;\Bigg[ \Cf \left(-1\right) + C^{tq}\left(2\right) + C^{t\bar{q}} \left(-2\right)
  \\
   & \hspace{-1cm} + C^{t\bar{t}} \left( \frac{8 (\ptop \ptopbar) \left(\mt^2 (p_q \ptopbar) (p_{\bar q} \ptop) + \mt^2 (p_{\bar q} \ptopbar) (p_q \ptop) - 2 (p_q \ptop) (p_{\bar q} \ptop) (\ptop \ptopbar) \right)}{\left(s_{t\bar{t}}-4\mt^2\right) s_{t\bar{t}} \left(\mt^2 (p_q p_{\bar q})-2 (p_q \ptop) (p_{\bar q} \ptop)\right)} \right) \Bigg],
\end{split}
\\[10pt]
\begin{split}
  \frac{ \delta_{\rm NP} \left [{\ptopbar}_{\perp}  \right ] }{{\ptopbar}_{\perp}} =& \frac{\alpha_s}{2 \pi }
  \frac{\pi \lambda}{\mt}\;\Bigg[ \Cf \left(-1\right) + C^{\bar{t} \bar{q}}\left(2\right) + C^{\bar{t} q} \left(-2\right)
  \\
   & \hspace{-1cm} + C^{t\bar{t}} \left( \frac{8 (\ptop \ptopbar) \left(\mt^2 (p_q \ptopbar) (p_{\bar q} \ptop) + \mt^2 (p_{\bar q} \ptopbar) (p_q \ptop) - 2 (p_q \ptopbar) (p_{\bar q} \ptopbar) (\ptop \ptopbar) \right)}{\left(s_{t\bar{t}}-4\mt^2\right) s_{t\bar{t}} \left(\mt^2 (p_q p_{\bar q})-2 (p_q \ptopbar) (p_{\bar q} \ptopbar)\right)} \right) \Bigg].
\end{split}
\end{align}

For the rapidity of the top and anti-top, we have that
\begin{align}
  \begin{split} 
    \delta_{\rm NP} \left [ y_t \right ]
    =& \frac{\alpha_s}{2 \pi } \frac{\pi \lambda}{\mt}\; \Bigg[ C^{t\bar{t}} \left(\frac{4 \mt^2 (\ptop \ptopbar)}{\left(s_{t\bar{t}}-4\mt^2\right) s_{t\bar{t}}} \left( \frac{p_q \ptopbar}{p_q \ptop} - \frac{p_{\bar q} \ptopbar}{p_{\bar q} \ptop} \right)\right)
\\    
    & \hspace{1.5cm} + C^{tq}\left(-\frac{\mt^2 \, (p_q p_{\bar q})}{(p_q \ptop) (p_{\bar q} \ptop)}\right) + C^{t\bar{q}}\left(-\frac{\mt^2 \, (p_q p_{\bar q})}{(p_q \ptop) (p_{\bar q} \ptop)}\right) \Bigg],
  \end{split} 
\\[10pt]
  \begin{split} 
    \delta_{\rm NP} \left [ y_{\bar{t}} \right ]
    =& \frac{\alpha_s}{2 \pi } \frac{\pi \lambda}{\mt}\; \Bigg[ C^{t\bar{t}} \left(\frac{4 \mt^2 (\ptop \ptopbar)}{\left(s_{t\bar{t}}-4\mt^2\right) s_{t\bar{t}}} \left( \frac{p_q \ptop}{p_q \ptopbar} - \frac{p_{\bar q} \ptop}{p_{\bar q} \ptopbar} \right)\right)
\\    
    & \hspace{1.5cm} + C^{\bar{t} q}\left(\frac{\mt^2 \, (p_q p_{\bar q})}{(p_q \ptopbar) (p_{\bar q} \ptopbar)}\right) + C^{\bar{t} \bar{q}}\left(\frac{\mt^2 \, (p_q p_{\bar q})}{(p_q \ptopbar) (p_{\bar q} \ptopbar)}\right) \Bigg].
  \end{split}
  \end{align}
  
The shift in the invariant mass of the $t\bar{t}$ pair reads,
\begin{align}
\begin{split}
\delta_{\rm NP} \left [ s_{t\bar{t}} \right ]
    =& \frac{\alpha_s}{2 \pi } \frac{\pi \lambda}{\mt} \bigg[ \Cf \left(-2s_{t\bar{t}}\right) + C^{t\bar{t}} \left(4 s_{t\bar{t}}-8\mt^2\right)
\\    
    & \hspace{1cm}  + C^{tq}\left(4 \ptop \ptopbar - 4\mt^2 \frac{\ptopbar p_q}{\ptop p_q}\right) + C^{t \bar{q}}\left(-4 \ptop \ptopbar + 4\mt^2 \frac{\ptopbar p_{\bar q}}{\ptop p_{\bar q}}\right)
\\    
    & \hspace{1cm}  + C^{\bar{t}q}\left(-4 \ptop \ptopbar + 4\mt^2 \frac{\ptop p_q}{\ptopbar p_q}\right) + C^{\bar{t} \bar{q}}\left(4 \ptop \ptopbar - 4\mt^2 \frac{\ptop p_{\bar q}}{\ptopbar p_{\bar q}}\right)   \bigg].
    \end{split}
\end{align}

For processes of type $e^+e^- \to t\bar{t} +X$ and the same observables, we can make use of the same expressions above but now need to adjust only for the colour coefficients as
   \begin{equation}
   \begin{split}
       &C^{tt} = C^{\bar t \bar t} = C^{t \bar t} = C_F, \hspace{2cm} C^{tq}= C^{ \bar{t} \bar{q} }=C^{\bar{t} q}= C^{ t \bar{q} } = 0.
   \end{split}
   \label{eq:coloure+e-}
   \end{equation}

\bibliographystyle{JHEP}
\bibliography{sintop}

\providecommand{\href}[2]{#2}\begingroup\raggedright\begin{thebibliography}{10}

\bibitem{ATLAS:2018mme}
{\scshape ATLAS} collaboration, \emph{{Observation of Higgs boson production in
  association with a top quark pair at the LHC with the ATLAS detector}},
  \href{https://doi.org/10.1016/j.physletb.2018.07.035}{\emph{Phys. Lett. B}
  {\bfseries 784} (2018) 173}
  [\href{https://arxiv.org/abs/1806.00425}{{\ttfamily 1806.00425}}].

\bibitem{ATLAS:2020ior}
{\scshape ATLAS} collaboration, \emph{{$CP$ Properties of Higgs Boson
  Interactions with Top Quarks in the $t\bar{t}H$ and $tH$ Processes Using $H
  \rightarrow \gamma\gamma$ with the ATLAS Detector}},
  \href{https://doi.org/10.1103/PhysRevLett.125.061802}{\emph{Phys. Rev. Lett.}
  {\bfseries 125} (2020) 061802}
  [\href{https://arxiv.org/abs/2004.04545}{{\ttfamily 2004.04545}}].

\bibitem{Mazzitelli:2021mmm}
J.~Mazzitelli, P.F.~Monni, P.~Nason, E.~Re, M.~Wiesemann and G.~Zanderighi,
  \emph{{Top-pair production at the LHC with MINNLO$_{PS}$}},
  \href{https://doi.org/10.1007/JHEP04(2022)079}{\emph{JHEP} {\bfseries 04}
  (2022) 079} [\href{https://arxiv.org/abs/2112.12135}{{\ttfamily
  2112.12135}}].

\bibitem{ATLAS:2020aln}
{\scshape ATLAS} collaboration, \emph{{Measurement of the $t\bar{t}$ production
  cross-section in the lepton+jets channel at $\sqrt{s}=13$ TeV with the ATLAS
  experiment}},
  \href{https://doi.org/10.1016/j.physletb.2020.135797}{\emph{Phys. Lett. B}
  {\bfseries 810} (2020) 135797}
  [\href{https://arxiv.org/abs/2006.13076}{{\ttfamily 2006.13076}}].

\bibitem{CMS:2020cga}
{\scshape CMS} collaboration, \emph{{Measurements of $\mathrm{t\bar{t}}H$
  Production and the CP Structure of the Yukawa Interaction between the Higgs
  Boson and Top Quark in the Diphoton Decay Channel}},
  \href{https://doi.org/10.1103/PhysRevLett.125.061801}{\emph{Phys. Rev. Lett.}
  {\bfseries 125} (2020) 061801}
  [\href{https://arxiv.org/abs/2003.10866}{{\ttfamily 2003.10866}}].

\bibitem{Czakon:2013goa}
M.~Czakon, P.~Fiedler and A.~Mitov, \emph{{Total Top-Quark Pair-Production
  Cross Section at Hadron Colliders Through $O(\alpha^4_S)$}},
  \href{https://doi.org/10.1103/PhysRevLett.110.252004}{\emph{Phys. Rev. Lett.}
  {\bfseries 110} (2013) 252004}
  [\href{https://arxiv.org/abs/1303.6254}{{\ttfamily 1303.6254}}].

\bibitem{Czakon:2016ckf}
M.~Czakon, P.~Fiedler, D.~Heymes and A.~Mitov, \emph{{NNLO QCD predictions for
  fully-differential top-quark pair production at the Tevatron}},
  \href{https://doi.org/10.1007/JHEP05(2016)034}{\emph{JHEP} {\bfseries 05}
  (2016) 034} [\href{https://arxiv.org/abs/1601.05375}{{\ttfamily
  1601.05375}}].

\bibitem{Catani:2020tko}
S.~Catani, S.~Devoto, M.~Grazzini, S.~Kallweit and J.~Mazzitelli,
  \emph{{Top-quark pair hadroproduction at NNLO: differential predictions with
  the $\overline{MS}$ mass}},
  \href{https://doi.org/10.1007/JHEP08(2020)027}{\emph{JHEP} {\bfseries 08}
  (2020) 027} [\href{https://arxiv.org/abs/2005.00557}{{\ttfamily
  2005.00557}}].

\bibitem{Kulesza:2020nfh}
A.~Kulesza, L.~Motyka, D.~Schwartl\"ander, T.~Stebel and V.~Theeuwes,
  \emph{{Associated top quark pair production with a heavy boson: differential
  cross sections at NLO+NNLL accuracy}},
  \href{https://doi.org/10.1140/epjc/s10052-020-7987-6}{\emph{Eur. Phys. J. C}
  {\bfseries 80} (2020) 428}
  [\href{https://arxiv.org/abs/2001.03031}{{\ttfamily 2001.03031}}].

\bibitem{Kidonakis:2022qvz}
N.~Kidonakis and A.~Tonero, \emph{{Higher-order corrections in
  tt\textasciimacron{}\ensuremath{\gamma} cross sections}},
  \href{https://doi.org/10.1103/PhysRevD.107.034013}{\emph{Phys. Rev. D}
  {\bfseries 107} (2023) 034013}
  [\href{https://arxiv.org/abs/2212.00096}{{\ttfamily 2212.00096}}].

\bibitem{Bevilacqua:2022ozv}
G.~Bevilacqua, M.~Lupattelli, D.~Stremmer and M.~Worek, \emph{{Study of
  additional jet activity in top quark pair production and decay at the LHC}},
  \href{https://doi.org/10.1103/PhysRevD.107.114027}{\emph{Phys. Rev. D}
  {\bfseries 107} (2023) 114027}
  [\href{https://arxiv.org/abs/2212.04722}{{\ttfamily 2212.04722}}].

\bibitem{Denner:2023eti}
A.~Denner, D.~Lombardi and G.~Pelliccioli, \emph{{Complete NLO corrections to
  off-shell $\text{t}\overline{\text{t}}\text{Z}$ production at the LHC}},
  \href{https://arxiv.org/abs/2306.13535}{{\ttfamily 2306.13535}}.

\bibitem{ATLAS:2014nxi}
{\scshape ATLAS} collaboration, \emph{{Measurement of the $t\bar{t}$ production
  cross-section using $e\mu $ events with b-tagged jets in pp collisions at
  $\sqrt{s}$ = 7 and 8 $\,\mathrm{TeV}$ with the ATLAS detector}},
  \href{https://doi.org/10.1140/epjc/s10052-016-4501-2}{\emph{Eur. Phys. J. C}
  {\bfseries 74} (2014) 3109}
  [\href{https://arxiv.org/abs/1406.5375}{{\ttfamily 1406.5375}}].

\bibitem{CMS:2016yys}
{\scshape CMS} collaboration, \emph{{Measurement of the t-tbar production cross
  section in the e-mu channel in proton-proton collisions at sqrt(s) = 7 and 8
  TeV}}, \href{https://doi.org/10.1007/JHEP08(2016)029}{\emph{JHEP} {\bfseries
  08} (2016) 029} [\href{https://arxiv.org/abs/1603.02303}{{\ttfamily
  1603.02303}}].

\bibitem{CMS:2018fks}
{\scshape CMS} collaboration, \emph{{Measurement of the
  $\mathrm{t}\overline{\mathrm{t}}$ production cross section, the top quark
  mass, and the strong coupling constant using dilepton events in pp collisions
  at $\sqrt{s} =$ 13 TeV}},
  \href{https://doi.org/10.1140/epjc/s10052-019-6863-8}{\emph{Eur. Phys. J. C}
  {\bfseries 79} (2019) 368}
  [\href{https://arxiv.org/abs/1812.10505}{{\ttfamily 1812.10505}}].

\bibitem{ATLAS:2019hau}
{\scshape ATLAS} collaboration, \emph{{Measurement of the $t\bar{t}$ production
  cross-section and lepton differential distributions in $e\mu $ dilepton
  events from $pp$ collisions at $\sqrt{s}=13\,\text {TeV}$ with the ATLAS
  detector}}, \href{https://doi.org/10.1140/epjc/s10052-020-7907-9}{\emph{Eur.
  Phys. J. C} {\bfseries 80} (2020) 528}
  [\href{https://arxiv.org/abs/1910.08819}{{\ttfamily 1910.08819}}].

\bibitem{ATLAS:2017dhr}
{\scshape ATLAS} collaboration, \emph{{Measurement of lepton differential
  distributions and the top quark mass in $t\bar{t}$ production in $pp$
  collisions at $\sqrt{s}=8$ TeV with the ATLAS detector}},
  \href{https://doi.org/10.1140/epjc/s10052-017-5349-9}{\emph{Eur. Phys. J. C}
  {\bfseries 77} (2017) 804}
  [\href{https://arxiv.org/abs/1709.09407}{{\ttfamily 1709.09407}}].

\bibitem{CMS:2019esx}
{\scshape CMS} collaboration, \emph{{Measurement of $\mathrm{t\bar t}$
  normalised multi-differential cross sections in pp collisions at $\sqrt s=13$
  TeV, and simultaneous determination of the strong coupling strength, top
  quark pole mass, and parton distribution functions}},
  \href{https://doi.org/10.1140/epjc/s10052-020-7917-7}{\emph{Eur. Phys. J. C}
  {\bfseries 80} (2020) 658}
  [\href{https://arxiv.org/abs/1904.05237}{{\ttfamily 1904.05237}}].

\bibitem{CMS:2014rml}
{\scshape CMS} collaboration, \emph{{Determination of the Top-Quark Pole Mass
  and Strong Coupling Constant from the $t \bar{t}$ Production Cross Section in
  $pp$ Collisions at $\sqrt{s}$ = 7 TeV}},
  \href{https://doi.org/10.1016/j.physletb.2013.12.009}{\emph{Phys. Lett. B}
  {\bfseries 728} (2014) 496}
  [\href{https://arxiv.org/abs/1307.1907}{{\ttfamily 1307.1907}}].

\bibitem{Juste:2013dsa}
A.~Juste, S.~Mantry, A.~Mitov, A.~Penin, P.~Skands, E.~Varnes et~al.,
  \emph{{Determination of the top quark mass circa 2013: methods, subtleties,
  perspectives}},
  \href{https://doi.org/10.1140/epjc/s10052-014-3119-5}{\emph{Eur. Phys. J. C}
  {\bfseries 74} (2014) 3119}
  [\href{https://arxiv.org/abs/1310.0799}{{\ttfamily 1310.0799}}].

\bibitem{Hoang:2014oea}
A.H.~Hoang, \emph{{The Top Mass: Interpretation and Theoretical
  Uncertainties}},  in \emph{{7th International Workshop on Top Quark
  Physics}}, 12, 2014 [\href{https://arxiv.org/abs/1412.3649}{{\ttfamily
  1412.3649}}].

\bibitem{Nason:2017cxd}
P.~Nason, \emph{{The Top Mass in Hadronic Collisions}},  in \emph{{From My Vast
  Repertoire ...}: {Guido Altarelli's Legacy}}, A.~Levy, S.~Forte and
  G.~Ridolfi, eds., pp.~123--151 (2019),
  \href{https://doi.org/10.1142/9789813238053_0008}{DOI}
  [\href{https://arxiv.org/abs/1712.02796}{{\ttfamily 1712.02796}}].

\bibitem{Azzi:2019yne}
P.~Azzi et~al., \emph{{Report from Working Group 1}: {Standard Model Physics at
  the HL-LHC and HE-LHC}},
  \href{https://doi.org/10.23731/CYRM-2019-007.1}{\emph{CERN Yellow Rep.
  Monogr.} {\bfseries 7} (2019) 1}
  [\href{https://arxiv.org/abs/1902.04070}{{\ttfamily 1902.04070}}].

\bibitem{Beneke:1998ui}
M.~Beneke, \emph{{Renormalons}},
  \href{https://doi.org/10.1016/S0370-1573(98)00130-6}{\emph{Phys. Rept.}
  {\bfseries 317} (1999) 1}
  [\href{https://arxiv.org/abs/hep-ph/9807443}{{\ttfamily hep-ph/9807443}}].

\bibitem{FerrarioRavasio:2018ubr}
S.~Ferrario~Ravasio, P.~Nason and C.~Oleari, \emph{{All-orders behaviour and
  renormalons in top-mass observables}},
  \href{https://doi.org/10.1007/JHEP01(2019)203}{\emph{JHEP} {\bfseries 01}
  (2019) 203} [\href{https://arxiv.org/abs/1810.10931}{{\ttfamily
  1810.10931}}].

\bibitem{Caola:2021kzt}
F.~Caola, S.~Ferrario~Ravasio, G.~Limatola, K.~Melnikov and P.~Nason, \emph{{On
  linear power corrections in certain collider observables}},
  \href{https://doi.org/10.1007/JHEP01(2022)093}{\emph{JHEP} {\bfseries 01}
  (2022) 093} [\href{https://arxiv.org/abs/2108.08897}{{\ttfamily
  2108.08897}}].

\bibitem{Makarov:2023ttq}
S.~Makarov, K.~Melnikov, P.~Nason and M.A.~Ozcelik, \emph{{Linear power
  corrections to single top production processes at the LHC}},
  \href{https://doi.org/10.1007/JHEP05(2023)153}{\emph{JHEP} {\bfseries 05}
  (2023) 153} [\href{https://arxiv.org/abs/2302.02729}{{\ttfamily
  2302.02729}}].

\bibitem{Dehnadi:2023msm}
B.~Dehnadi, A.H.~Hoang, O.L.~Jin and V.~Mateu, \emph{{Top Quark Mass
  Calibration for Monte Carlo Event Generators -- An Update}},
  \href{https://arxiv.org/abs/2309.00547}{{\ttfamily 2309.00547}}.

\bibitem{Bachu:2020nqn}
B.~Bachu, A.H.~Hoang, V.~Mateu, A.~Pathak and I.W.~Stewart, \emph{{Boosted top
  quarks in the peak region with NL3L resummation}},
  \href{https://doi.org/10.1103/PhysRevD.104.014026}{\emph{Phys. Rev. D}
  {\bfseries 104} (2021) 014026}
  [\href{https://arxiv.org/abs/2012.12304}{{\ttfamily 2012.12304}}].

\bibitem{Hoang:2017kmk}
A.H.~Hoang, S.~Mantry, A.~Pathak and I.W.~Stewart, \emph{{Extracting a Short
  Distance Top Mass with Light Grooming}},
  \href{https://doi.org/10.1103/PhysRevD.100.074021}{\emph{Phys. Rev. D}
  {\bfseries 100} (2019) 074021}
  [\href{https://arxiv.org/abs/1708.02586}{{\ttfamily 1708.02586}}].

\bibitem{Hoang:2019ceu}
A.H.~Hoang, S.~Mantry, A.~Pathak and I.W.~Stewart, \emph{{Nonperturbative
  Corrections to Soft Drop Jet Mass}},
  \href{https://doi.org/10.1007/JHEP12(2019)002}{\emph{JHEP} {\bfseries 12}
  (2019) 002} [\href{https://arxiv.org/abs/1906.11843}{{\ttfamily
  1906.11843}}].

\bibitem{Low:1958sn}
F.E.~Low, \emph{{Bremsstrahlung of very low-energy quanta in elementary
  particle collisions}},
  \href{https://doi.org/10.1103/PhysRev.110.974}{\emph{Phys. Rev.} {\bfseries
  110} (1958) 974}.

\bibitem{Burnett:1967km}
T.H.~Burnett and N.M.~Kroll, \emph{{Extension of the low soft photon theorem}},
  \href{https://doi.org/10.1103/PhysRevLett.20.86}{\emph{Phys. Rev. Lett.}
  {\bfseries 20} (1968) 86}.

\bibitem{Engel:2021ccn}
T.~Engel, A.~Signer and Y.~Ulrich, \emph{{Universal structure of radiative QED
  amplitudes at one loop}},
  \href{https://doi.org/10.1007/JHEP04(2022)097}{\emph{JHEP} {\bfseries 04}
  (2022) 097} [\href{https://arxiv.org/abs/2112.07570}{{\ttfamily
  2112.07570}}].

\bibitem{Lebiedowicz:2021byo}
P.~Lebiedowicz, O.~Nachtmann and A.~Szczurek, \emph{{High-energy
  \ensuremath{\pi}\ensuremath{\pi} scattering without and with photon
  radiation}}, \href{https://doi.org/10.1103/PhysRevD.105.014022}{\emph{Phys.
  Rev. D} {\bfseries 105} (2022) 014022}
  [\href{https://arxiv.org/abs/2107.10829}{{\ttfamily 2107.10829}}].

\bibitem{Lebiedowicz:2023mlz}
P.~Lebiedowicz, O.~Nachtmann and A.~Szczurek, \emph{{Soft-photon theorem for
  pion-proton elastic scattering revisited}},
  \href{https://arxiv.org/abs/2307.12673}{{\ttfamily 2307.12673}}.

\bibitem{Lebiedowicz:2023ell}
P.~Lebiedowicz, O.~Nachtmann and A.~Szczurek, \emph{{Soft-Photon Theorem for
  Pion-Proton Scattering: Next to Leading Term}},
  \href{https://arxiv.org/abs/2307.13291}{{\ttfamily 2307.13291}}.

\bibitem{Peter:1996ig}
M.~Peter, \emph{{The Static quark - anti-quark potential in QCD to three
  loops}}, \href{https://doi.org/10.1103/PhysRevLett.78.602}{\emph{Phys. Rev.
  Lett.} {\bfseries 78} (1997) 602}
  [\href{https://arxiv.org/abs/hep-ph/9610209}{{\ttfamily hep-ph/9610209}}].

\bibitem{Schroder:1998vy}
Y.~Schroder, \emph{{The Static potential in QCD to two loops}},
  \href{https://doi.org/10.1016/S0370-2693(99)00010-6}{\emph{Phys. Lett. B}
  {\bfseries 447} (1999) 321}
  [\href{https://arxiv.org/abs/hep-ph/9812205}{{\ttfamily hep-ph/9812205}}].

\bibitem{tHooft:1973mfk}
G.~'t~Hooft, \emph{{Dimensional regularization and the renormalization group}},
  \href{https://doi.org/10.1016/0550-3213(73)90376-3}{\emph{Nucl. Phys. B}
  {\bfseries 61} (1973) 455}.

\bibitem{Czarnecki:1997sz}
A.~Czarnecki, K.~Melnikov and N.~Uraltsev, \emph{{Non-Abelian dipole radiation
  and the heavy quark expansion}},
  \href{https://doi.org/10.1103/PhysRevLett.80.3189}{\emph{Phys. Rev. Lett.}
  {\bfseries 80} (1998) 3189}
  [\href{https://arxiv.org/abs/hep-ph/9708372}{{\ttfamily hep-ph/9708372}}].

\bibitem{Beneke:1998rk}
M.~Beneke, \emph{{A Quark mass definition adequate for threshold problems}},
  \href{https://doi.org/10.1016/S0370-2693(98)00741-2}{\emph{Phys. Lett. B}
  {\bfseries 434} (1998) 115}
  [\href{https://arxiv.org/abs/hep-ph/9804241}{{\ttfamily hep-ph/9804241}}].

\bibitem{Hoang:1999ye}
A.H.~Hoang, \emph{{1S and MS-bar bottom quark masses from Upsilon sum rules}},
  \href{https://doi.org/10.1103/PhysRevD.61.034005}{\emph{Phys. Rev. D}
  {\bfseries 61} (2000) 034005}
  [\href{https://arxiv.org/abs/hep-ph/9905550}{{\ttfamily hep-ph/9905550}}].

\bibitem{Pineda:2001zq}
A.~Pineda, \emph{{Determination of the bottom quark mass from the Upsilon(1S)
  system}}, \href{https://doi.org/10.1088/1126-6708/2001/06/022}{\emph{JHEP}
  {\bfseries 06} (2001) 022}
  [\href{https://arxiv.org/abs/hep-ph/0105008}{{\ttfamily hep-ph/0105008}}].

\bibitem{Hoang:2008yj}
A.H.~Hoang, A.~Jain, I.~Scimemi and I.W.~Stewart, \emph{{Infrared
  Renormalization Group Flow for Heavy Quark Masses}},
  \href{https://doi.org/10.1103/PhysRevLett.101.151602}{\emph{Phys. Rev. Lett.}
  {\bfseries 101} (2008) 151602}
  [\href{https://arxiv.org/abs/0803.4214}{{\ttfamily 0803.4214}}].

\bibitem{FerrarioRavasio:2020guj}
S.~Ferrario~Ravasio, G.~Limatola and P.~Nason, \emph{{Infrared renormalons in
  kinematic distributions for hadron collider processes}},
  \href{https://doi.org/10.1007/JHEP06(2021)018}{\emph{JHEP} {\bfseries 06}
  (2021) 018} [\href{https://arxiv.org/abs/2011.14114}{{\ttfamily
  2011.14114}}].

\bibitem{Beneke:2016cbu}
M.~Beneke, P.~Marquard, P.~Nason and M.~Steinhauser, \emph{{On the ultimate
  uncertainty of the top quark pole mass}},
  \href{https://doi.org/10.1016/j.physletb.2017.10.054}{\emph{Phys. Lett. B}
  {\bfseries 775} (2017) 63}
  [\href{https://arxiv.org/abs/1605.03609}{{\ttfamily 1605.03609}}].

\bibitem{Hoang:2017btd}
A.H.~Hoang, C.~Lepenik and M.~Preisser, \emph{{On the Light Massive Flavor
  Dependence of the Large Order Asymptotic Behavior and the Ambiguity of the
  Pole Mass}}, \href{https://doi.org/10.1007/JHEP09(2017)099}{\emph{JHEP}
  {\bfseries 09} (2017) 099}
  [\href{https://arxiv.org/abs/1706.08526}{{\ttfamily 1706.08526}}].

\bibitem{Schwienhorst:2022yqu}
K.~Agashe et~al., \emph{{Report of the Topical Group on Top quark physics and
  heavy flavor production for Snowmass 2021}},
  \href{https://arxiv.org/abs/2209.11267}{{\ttfamily 2209.11267}}.

\bibitem{NNPDF:2017mvq}
{\scshape NNPDF} collaboration, \emph{{Parton distributions from high-precision
  collider data}},
  \href{https://doi.org/10.1140/epjc/s10052-017-5199-5}{\emph{Eur. Phys. J. C}
  {\bfseries 77} (2017) 663}
  [\href{https://arxiv.org/abs/1706.00428}{{\ttfamily 1706.00428}}].

\end{thebibliography}\endgroup

\end{document}